\documentclass[twocolumn, 10pt]{article}

\usepackage[margin=0.8in]{geometry}
\usepackage{times}
\usepackage{graphicx}
\usepackage{caption}
\usepackage{booktabs}
\usepackage{multirow}
\usepackage{stfloats}
\usepackage{amsmath, amssymb}
\usepackage{bm}
\usepackage{authblk}

\usepackage{cite}
\usepackage{url}
\usepackage{hyperref}
\hypersetup{
	colorlinks=true,
	linkcolor=blue,
	filecolor=magenta,
	urlcolor=blue,
	citecolor=blue,
}

\usepackage{titlesec}
\titleformat{\section}{\large\bfseries}{\thesection}{0.5em}{}
\titleformat{\subsection}{\normalsize\bfseries}{\thesubsection}{0.5em}{}
\titlespacing*{\section}{0pt}{2.5ex}{1.5ex}
\titlespacing*{\subsection}{0pt}{2ex}{1ex}

\title{\textbf{Spectral analysis of protein backbone geometry reveals abrupt helix--coil boundaries}}

\author[1,2,3]{Yiquan Wang}

\affil[1]{College of Mathematics and System Science, Xinjiang University, Urumqi, Xinjiang, China}
\affil[2]{Xinjiang Key Laboratory of Biological Resources and Genetic Engineering, College of Life Science and Technology, Xinjiang University, Urumqi, Xinjiang, China}
\affil[3]{Shenzhen X-Institute, Shenzhen, China}

\date{}

\begin{document}

\maketitle
\renewcommand*{\thefootnote}{\fnsymbol{footnote}}
\footnotetext[1]{Corresponding author:
    Yiquan Wang (\href{mailto:ethan@stu.xju.edu.cn}{ethan@stu.xju.edu.cn})}
\renewcommand*{\thefootnote}{\arabic{footnote}}

\begin{abstract}
	The boundaries of cooperative helix--coil transitions influence protein allostery and conformational dynamics, yet the persistent one-to-two-residue ambiguity in their assignment remains poorly characterized. We apply the discrete Hasimoto map to translate three-dimensional C$_\alpha$ backbone geometry into a one-dimensional discrete nonlinear Schr\"{o}dinger effective potential and analyze its spatial-frequency structure. Helical segments appear as near-integrable, low-entropy states whose spectral power concentrates at the zero-frequency mode, whereas coil regions show broadband fluctuations. A pointwise integrability residual and a windowed spectral entropy separate the two phases with ROC AUC values of 0.783 and 0.715, and their combination reaches 0.803, while combining the residual instead with a low-frequency energy ratio reaches 0.815. Across 1\,986 proteins and 19\,148 of 21\,107 fitted helix--coil boundaries the transition is abrupt, with a median sigmoid width of 0.145 residues that measures the steepness of a single-step discrete jump rather than a literal sub-residue distance; the transition is directionally asymmetric, with helix exits sharper than entries. Across the full dataset every C$_\alpha$ geometry-based assignment, including DSSP-calibrated P-SEA and both spectral probes, loses agreement with the DSSP hydrogen-bond reference most acutely at these boundaries, indicating that the assignment ambiguity is a general feature of C$_\alpha$ geometry rather than any single algorithm. The windowed spectral probe is subject to a Gabor resolution limit and is therefore outperformed by the pointwise probe, which attains the lattice-limited resolution.

\end{abstract}
\vspace{0.5em}


	\section{Introduction}
	\label{sec:intro}
	
	\begin{figure*}[t]
		\centering
		\includegraphics[width=\textwidth]{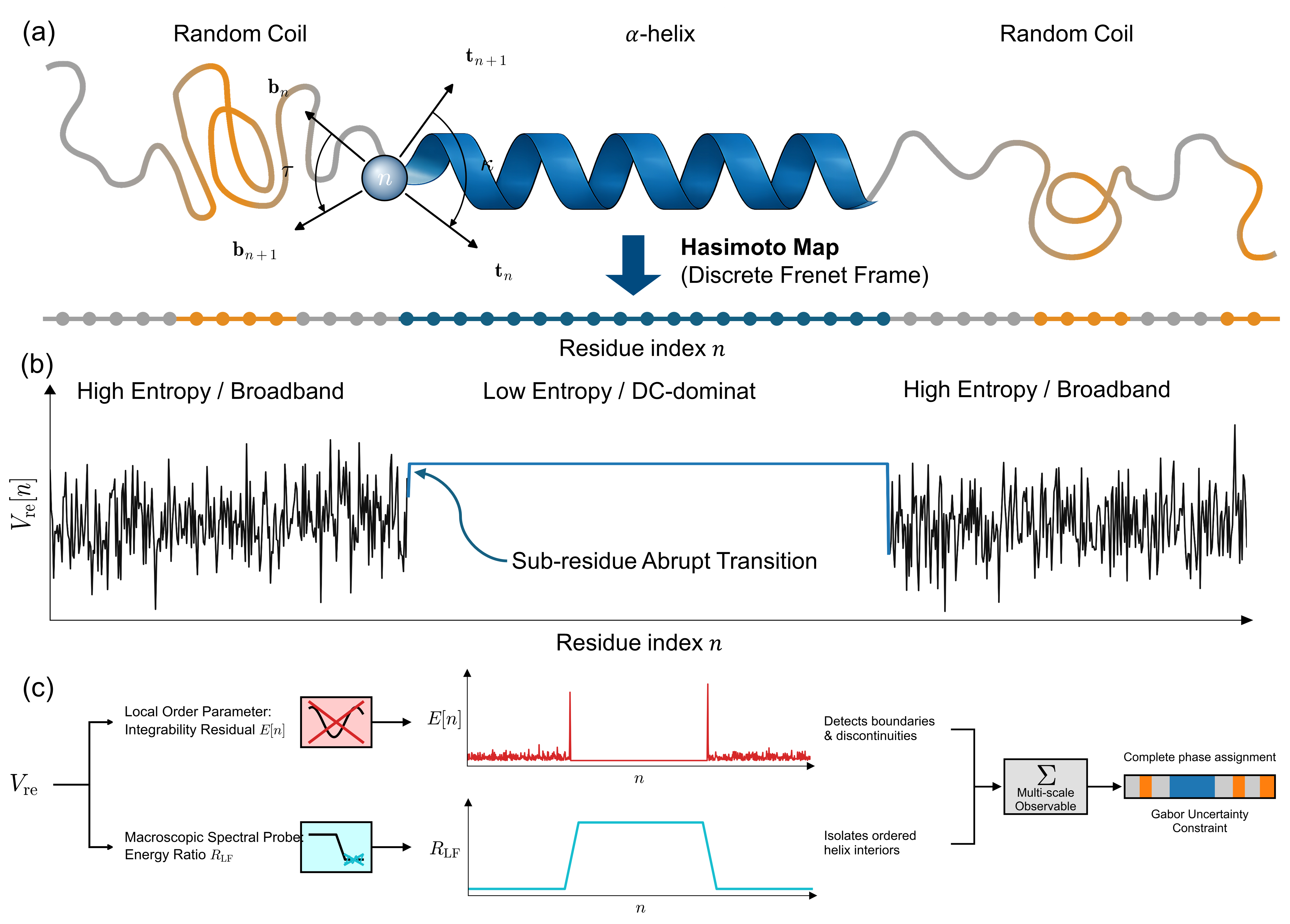}
		\caption{Multi-scale structural characterization of protein helix--coil phase boundaries via the discrete Hasimoto map.
			(a) Geometric projection. The 3D backbone geometry (bond-angle curvature $\kappa$ and torsion $\tau$) is projected onto a 1D lattice via the discrete Frenet frame.
			(b) Order--disorder dichotomy. The Hasimoto effective potential $V_{\mathrm{re}}[n]$ maps structurally ordered helical states to flat, negative plateaus characteristic of near-integrable soliton-like regimes. Disordered coil regions appear as large-amplitude conformational fluctuations. The phase boundaries between these structural states exhibit sharp, single-step transitions.
			(c) Multi-scale characterization. Structural phase-boundary detection exploits two complementary observables at different spatial scales. The integrability residual $E[n]$ serves as a pointwise local order parameter, vanishing within the helical interior while sensitively detecting boundary discontinuities (red spikes). The low-frequency energy ratio $R_{\mathrm{LF}}$ serves as a macroscopic spectral probe that isolates the uniform spectral core of the ordered helix. Combining these multi-scale observables captures the geometric context across the resolution regimes of the pointwise and windowed spectral probes.}
		\label{fig:concept}
	\end{figure*}
	
	The protein backbone constitutes a semiflexible heteropolymer chain \cite{witzky2025heteropolymer, kravikass2024isolated} whose secondary structures ($\alpha$-helices, $\beta$-sheets, and coils) arise from a competition between hydrogen-bond enthalpies and conformational entropy. This competition operates at the scale of a few times the thermal energy $k_B T$, and the helix--coil transition represents a canonical one-dimensional highly cooperative order--disorder transition \cite{zimm1959theory}, often described as a pseudo-phase transition in finite-size biopolymers owing to the absence of true long-range order in one-dimensional systems with short-range interactions \cite{kittel1969phase, poland1966occurrence}. Recent topological analysis of configuration-space isopotential hypersurfaces supports its thermodynamic phase-transition character in the topological sense \cite{dicairano2022topological}. Precisely delineating the geometric boundaries of these structural phases is relevant to macromolecular folding \cite{ji2010role}, allosteric signal propagation, and conformational dynamics, since allosteric transitions often involve one-to-two-residue displacements of helical boundaries \cite{laskowski2009structural, weis2018molecular} and thermal helix fraying induces analogous fluctuations smaller than one residue spacing at sub-nanosecond timescales \cite{daggett1992molecular}. At each C$_\alpha$ position $n$ along the backbone, the bond-angle curvature $\kappa[n]$ and dihedral torsion $\tau[n]$ characterize the three-dimensional space curve; in $\alpha$-helical segments both quantities adopt nearly constant values dictated by regular hydrogen-bonding patterns, while in coil regions they fluctuate widely. Despite these intrinsic conformational fluctuations, secondary-structure assignment algorithms, whether based on hydrogen-bond energetics \cite{kabsch1983dictionary, frishman1995knowledge}, backbone geometry \cite{labesse1997p, martin2005protein, khalife2021secondary}, sequence-based prediction \cite{rost1993prediction}, or end-to-end deep learning \cite{jumper2021highly}, systematically disagree by one to two residues at helix--coil interfaces. Whether this persistent ambiguity reflects differing algorithmic definitions or the limited resolving power of the geometric probes used to detect the boundary remains unresolved.

	Differential-geometric and spectral approaches have probed backbone structure from complementary angles. Curvature--torsion analyses of protein space curves \cite{rackovsky1984differential, louie1983differential, feoli2005functionals} and signal-processing methods that extract periodicities from one-dimensional physicochemical profiles \cite{eisenberg1982hydrophobic, eisenberg1984hydrophobic, cornette1987hydrophobicity, cheng2012detecting, pando2009detection, benson2012wavelet, veljkovic1972simple, veljkovic1985possible, cosic2002macromolecular, wang2025signal} have each illuminated aspects of secondary-structure regularity, yet these methods predominantly operate on scalar sequences rather than the full three-dimensional backbone geometry, and none provides a physical mechanism governing boundary sharpness.

	A framework for connecting three-dimensional backbone geometry to one-dimensional physical fields originates in vortex filament dynamics. The Hasimoto map \cite{hasimoto1972soliton} combines curvature and torsion into a complex scalar field governed by the nonlinear Schr\"{o}dinger equation, whose soliton solutions motivated the topological-soliton description of protein folds \cite{chernodub2010topological, davydov1977solitons}. The discrete formulation of this map \cite{danielsson2010gauge, hu2011discrete} yields a discrete nonlinear Schr\"{o}dinger equation (DNLS), under which $\alpha$-helices emerge as soliton excitations with sub-angstrom parametric accuracy across thousands of proteins \cite{molkenthin2011discrete, krokhotin2012soliton}. Recent extensions have modeled topological transitions via Arnold perestroikas \cite{begun2025local} and connected soliton dynamics to protein knot formation \cite{begun2021topology}, while independent work has identified secondary structures through discrete Frenet criteria \cite{prados2025frenet}. These studies establish the descriptive power of the DNLS soliton picture but treat known structures as fitting targets (the inverse problem) without exploiting the spectral content of the effective potential for forward physical analysis. Recent work \cite{wang2026structural} derived an exact closed-form decomposition of the DNLS effective potential and established that the Hasimoto map is a lossless kinematic identity rather than a dynamical equation, so that the real potential $V_{\text{re}}$ represents pure backbone geometry available for direct spectral analysis. Subsequent research \cite{wang2026piecewise} introduced the concept of piecewise integrability, showing that helical segments correspond to near-integrable islands where the analytic dispersion relation reconstructs backbone coordinates with sub-angstrom accuracy (median RMSD $= 0.77$\,\AA), and that the integrability residual $E[n]$ maps the geometric boundaries of these islands. Whether this boundary is sharp or gradual when read from the spectral structure of the effective potential, and how the resolving power of windowed versus pointwise spectral probes limits its measurement, remain unquantified.
	
	The discrete Hasimoto map provides the mathematical framework for projecting the three-dimensional backbone onto a one-dimensional physical field amenable to spectral analysis. It combines curvature and torsion into the complex scalar field $\psi[n] = \kappa[n]\,\exp(i\sum_{k=1}^{n} \tau[k])$, which evolves according to a discrete nonlinear Schr\"{o}dinger (DNLS) equation \cite{orland2002rna} through an effective potential $V_{\mathrm{eff}}[n] = V_{\mathrm{re}}[n] + i V_{\mathrm{im}}[n]$ that encodes the deviation of the backbone from an ideal integrable curve. The real part $V_{\mathrm{re}}[n]$ admits an exact analytic decomposition \cite{wang2026structural},
	\begin{equation}
		\label{eq:Vre}
		V_{\mathrm{re}}[n] = \beta^{+} r^{+}[n]\cos\tau[n{+}1] + \beta^{-} r^{-}[n]\cos\tau[n] - (\beta^{+}+\beta^{-}),
	\end{equation}
	where $r^{\pm}[n] = \kappa[n{\pm}1]/\kappa[n]$ are dimensionless nearest-neighbor curvature ratios and $\beta^{\pm} = 1/|\mathbf{t}[n{\pm}1]|$ are inverse virtual-bond lengths (dimension \AA$^{-1}$); since all remaining factors are dimensionless, $V_{\mathrm{re}}$ inherits the dimension \AA$^{-1}$. The imaginary part takes the form
	\begin{equation}
		\label{eq:Vim}
		V_{\mathrm{im}}[n] = \beta^+ r^+[n] \sin\tau[n{+}1]
		                    - \beta^- r^-[n] \sin\tau[n],
	\end{equation}
	which encodes the local backbone chirality through the odd symmetry of $\sin\tau$. Under a torsion-sign flip $\tau[n] \to -\tau[n]$, $V_{\mathrm{re}}$ remains invariant while $V_{\mathrm{im}}$ changes sign, generating a $2^L$-fold chiral degeneracy if the imaginary part is discarded \cite{wang2026structural}. In geometrically uniform helical segments, where $\tau[n] \approx \tau_0$ and $r^{\pm} \approx 1$, the two terms cancel, yielding $V_{\mathrm{im}} \approx 0$; at helix--coil boundaries, the torsion mismatch between adjacent sites produces a nonzero imaginary potential. Since our STFT analysis exploits the spectral contrast between the near-constant $V_{\mathrm{re}}$ plateaus of helical segments (narrow-band, DC-dominated) and the broadband fluctuations of coil regions, and $V_{\mathrm{im}}$ vanishes throughout helical interiors regardless of helix type or handedness, the real part alone provides the complete spectral signature required for helix--coil discrimination. We therefore focus exclusively on $V_{\mathrm{re}}$ in the present work. As shown in Ref.~\cite{wang2026structural}, this map constitutes a kinematic identity for proteins, and the integrability residual
	\begin{equation}
		\label{eq:E_def}
		E[n] = \left|\cos\tau[n] - \left(1 + \frac{V_{\mathrm{re}}[n]}{2\bar{\beta}}\right)\right|,
	\end{equation}
	measures the departure of local geometry from the DNLS dispersion relation \cite{wang2026structural}. Under a uniform-curvature approximation ($r^{\pm} \approx 1$), $E[n]$ reduces to $E[n] \approx \tfrac{1}{2}|\cos\tau[n] - \cos\tau[n{+}1]|$, directly quantifying how abruptly the backbone torsion geometry changes between adjacent residues \cite{wang2026piecewise}. This expression makes explicit that $E[n] \to 0$ in geometrically uniform helical regions where the backbone satisfies the DNLS dispersion relation at every site, while $E[n]$ exhibits sharp peaks at helix--coil boundaries where torsion non-uniformity disrupts the local integrability. This property enables identification of piecewise integrable islands for highly accurate three-dimensional analytic reconstruction. These spatial-domain results establish the Hasimoto potential as a physically grounded structural order parameter, and in this study we analyze the spatial-frequency structure of $V_{\mathrm{re}}[n]$ to investigate the physical mechanisms governing the sharpness and spatial extent of helix--coil phase boundaries.
	
	In the spatial-frequency domain, helical and coil geometries occupy distinct spectral regimes. Helical segments appear as near-integrable, low-entropy soliton-like states whose spectral power concentrates at the zero-frequency mode, reflecting their geometrically uniform character, whereas coil regions exhibit broadband conformational fluctuations distributed across all spatial frequencies. The helix--coil phase boundaries are abrupt, and across 1\,986 proteins we measure a distribution of geometric transition widths together with a directional asymmetry in which helix exits are sharper than entries (Sec.~\ref{sec:boundary_sharpness}).
	
	We further show that the windowed spectral probe is subject to a Gabor resolution limit: expanding the observation window improves frequency resolution but progressively blurs the boundary, so that phase-boundary distinguishability is maximized at the narrowest feasible window \cite{gabor1946electrical}. This trade-off is a property of the windowed spectral method rather than of the backbone geometry itself, and it explains why the pointwise integrability residual $E[n]$ \cite{wang2026structural, wang2026piecewise}, which carries no observation window, resolves boundaries more sharply than the windowed spectral entropy.
	
	Because the sharpness of these boundaries governs where flexibility localizes along the backbone, with consequences for folding, allosteric signaling, and conformational dynamics, we analyze the effective potential in the spatial-frequency domain, which exposes a duality between the local geometric order of helical interiors and the macroscopic spectral order of the chain. This physical duality motivates a multi-scale structural characterization framework (Fig.~\ref{fig:concept}). The discrete Hasimoto map translates three-dimensional backbone geometry into a one-dimensional effective potential (Fig.~\ref{fig:concept}a), mapping structurally ordered helices to near-integrable soliton-like states and disordered coils to broadband conformational fluctuations (Fig.~\ref{fig:concept}b). The pointwise integrability residual $E[n]$ serves as a local order parameter that captures torsion discontinuities at structural phase boundaries, while a complementary low-frequency energy ratio $R_{\mathrm{LF}}$ serves as a macroscopic spectral probe that measures the spectral uniformity of helical interiors (Fig.~\ref{fig:concept}c). Combining these multi-scale observables reconciles local boundary precision with global topological robustness and improves structural-state separability. This framework suggests that broadband, high-entropy geometries may correspond to the conformational flexibility associated with allosteric transitions and protein interactions, a hypothesis that motivates future validation against experimental dynamics data.
	
	\section{Methods}
	\label{sec:methods}
	
	We constructed two datasets from the RCSB Protein Data Bank \cite{berman2000protein}. The main dataset is derived from the CullPDB non-redundant list \cite{wang2003pisces} (sequence identity $\leq 25\%$, resolution $\leq 2.0$~\AA, X-ray structures only, chain length 40--300 residues, downloaded February 2026), which provides 2\,004 unique chains. Each chain was subjected to four quality filters, requiring at least 7 C$_\alpha$ atoms (the minimum required to produce at least one $V_{\mathrm{re}}$ value after Frenet-frame trimming), all C$_\alpha$--C$_\alpha$ virtual bond lengths within the range 1.0--10.0~\AA\ (to exclude chains with missing residues or coordinate errors; the standard C$_\alpha$--C$_\alpha$ distance is approximately 3.8~\AA), no vanishing bond-angle curvature ($\kappa > 10^{-15}$~rad at every position, to prevent division by zero in the curvature ratios $r^{\pm}$), and successful DSSP secondary structure calculation. After applying these filters, the final set comprises 1,986 chains (320,453 aligned residues), of which 852 overlap with the original 856-protein dataset of Ref.~\cite{wang2026structural}. The helical peptide dataset was assembled by selecting all main-dataset chains with DSSP helix fraction $\geq 70\%$ and augmenting with the 50 high-helicity short peptides from Ref.~\cite{wang2026piecewise} (20--50 residues, helix fraction $> 85\%$; 49 of 50 passed the quality filters above), yielding 251 chains (mean helix fraction 83.4\%, length range 22--298). Table~\ref{tab:dataset} summarizes both datasets. Secondary structure labels were computed using DSSP (\texttt{mkdssp}) \cite{kabsch1983dictionary, joosten2010series}, which identifies hydrogen bonds from backbone N--H and C=O electrostatic interaction energies (threshold $< -0.5$ kcal/mol) and assigns one of eight states based on the resulting hydrogen-bonding patterns. For the primary analysis, these eight states were reduced to three classes, namely helix (H, G, I $\to$~H, encompassing $\alpha$-helix, $3_{10}$-helix, and $\pi$-helix), sheet (E, B $\to$~E, encompassing $\beta$-strand and $\beta$-bridge), and coil (T, S, blank $\to$~C, encompassing turn, bend, and unstructured regions). This three-state reduction groups all helical subtypes together because the Hasimoto effective potential does not distinguish between them at the level of $V_{\mathrm{re}}[n]$; all three helix types produce similar near-constant negative plateaus. For the structural correlate analysis in Sec.~\ref{sec:boundary_sharpness}, the original eight-state labels were also retained to distinguish $\alpha$-helix (H) from $3_{10}$-helix (G) and $\pi$-helix (I) at helix--coil boundaries. In the main dataset, the three-state distribution is 36.4\% helix, 25.6\% sheet, and 38.0\% coil; within the helix class, $\alpha$-helix accounts for 31.8\%, $3_{10}$-helix 4.1\%, and $\pi$-helix 0.5\% of all residues. Index alignment between the Frenet-frame geometric quantities and DSSP labels used \texttt{dssp}$[2{:}N{-}2]$, trimming two residues from each end to match the index range of $V_{\mathrm{re}}[n]$ and $E[n]$.
	
	\begin{table}[htb]
		\caption{\label{tab:dataset}Summary statistics of the two datasets. \textit{Aligned residues} refers to positions for which $V_{\mathrm{re}}$, $E$, and DSSP labels are simultaneously available after Frenet-frame trimming.}
			\begin{tabular}{lcc}
				\toprule
				& Main dataset & Helical peptides \\
				\midrule
				Chains & 1\,986 & 251 \\
				Aligned residues & 320\,453 & --- \\
				Length range & 40--300 & 22--298 \\
				Mean length & 165.4 & 106.5 \\
				Helix fraction & 36.4\% & 83.4\% (mean) \\
				Sheet fraction & 25.6\% & --- \\
				Coil fraction & 38.0\% & --- \\
				Overlap with Ref.~\cite{wang2026structural} & 852 chains & --- \\
				Overlap with Ref.~\cite{wang2026piecewise} & --- & 49 chains \\
				\bottomrule
			\end{tabular}
	\end{table}
	
	For each chain, the discrete Frenet--Serret frame was computed from the C$_\alpha$ trace as described in \cite{wang2026structural, danielsson2010gauge}. Given $N$ C$_\alpha$ coordinates $\mathbf{r}[n]$ ($n = 1, \ldots, N$), the unit tangent vectors $\mathbf{t}[n] = \Delta\mathbf{r}[n] / |\Delta\mathbf{r}[n]|$ were formed from successive differences $\Delta\mathbf{r}[n] = \mathbf{r}[n{+}1] - \mathbf{r}[n]$. The bond-angle curvature was computed as $\kappa[n] = \arccos(\mathbf{t}[n] \cdot \mathbf{t}[n{+}1])$ ($N{-}2$ values), and the dihedral torsion as $\tau[n] = \mathrm{atan2}((\hat{\mathbf{b}}[n] \times \hat{\mathbf{b}}[n{+}1]) \cdot \mathbf{t}[n{+}1],\; \hat{\mathbf{b}}[n] \cdot \hat{\mathbf{b}}[n{+}1])$ ($N{-}3$ values), where $\hat{\mathbf{b}}$ is the unit binormal vector obtained by normalizing $\mathbf{t}[n] \times \mathbf{t}[n{+}1]$. The DNLS effective potential components $V_{\mathrm{re}}[n]$ and $V_{\mathrm{im}}[n]$ were then computed from Eqs.~(\ref{eq:Vre})--(\ref{eq:Vim}), and the integrability residual $E[n]$ from Eq.~(\ref{eq:E_def}), yielding $L = N{-}4$ aligned values per chain (after trimming two residues from each end of the Frenet frame). The bond-length parameters $\beta^{\pm}[n] = 1/b[n{\pm}1]$ were computed from actual C$_\alpha$--C$_\alpha$ distances with no uniform-bond-length assumption (real structures exhibit small deviations from the standard value noted above, owing to peptide bond geometry and coordinate uncertainty), and the global average $\bar{\beta} = \tfrac{1}{2} (\langle\beta^{+}\rangle + \langle\beta^{-}\rangle)$ was computed per chain. The curvature ratios $r^{\pm}[n] = \kappa[n{\pm}1] / \kappa[n]$ quantify the local uniformity of bending: in ideal helices $r^{\pm} \approx 1$, while in coil regions $r^{\pm}$ can deviate substantially from unity. These spatial-domain geometric quantities provide the input signal for the frequency-domain analysis described next.

	The protein backbone can be viewed as a one-dimensional lattice indexed by residue number $n$, and the effective potential $V_{\mathrm{re}}[n]$ as a scalar field defined on this lattice. In ordered helical segments, $V_{\mathrm{re}}[n]$ is approximately translationally invariant (constant along $n$), reflecting the spatial periodicity imposed by regular hydrogen-bonding geometry. The helix--coil transition breaks this local translational symmetry, and coil regions exhibit rapid, aperiodic fluctuations in $V_{\mathrm{re}}$. The short-time Fourier transform (STFT) probes this local symmetry breaking by decomposing $V_{\mathrm{re}}[n]$ into spatial-frequency components within a sliding window. Because the independent variable is the residue index rather than time, the resulting transform yields spatial frequencies that measure structural periodicity along the backbone. The zero-frequency (DC) component corresponds to a segment of constant $V_{\mathrm{re}}$, implying uniform curvature $\kappa$ and torsion $\tau$; in three dimensions, such a segment traces a regular helix of fixed radius and pitch. Near-integrable helical segments concentrate spectral power at this DC mode, while coil regions distribute power broadly across all spatial frequencies; the local power spectrum therefore quantifies the degree of local structural order.

	The short-time Fourier transform of $V_{\mathrm{re}}[n]$ was computed with the SciPy library \cite{virtanen2020scipy} using a Gaussian window of standard deviation $\sigma$ (in residues). The Gaussian window was chosen because it achieves the minimum time-frequency uncertainty product $\Delta t \cdot \Delta \omega = 1/2$, providing the best simultaneous localization in both domains \cite{gabor1946electrical}. This property is essential for protein backbone signals, where secondary structure segments can be as short as 3--4 residues ($3_{10}$-helices) and the transitions between them occur within 1--2 residues. The window length was set to $M = \mathrm{round}(6\sigma)$ (forced to the nearest odd integer), which captures $99.7\%$ of the Gaussian weight, and the overlap was set to $M{-}1$ (single-sample stride) to produce one spectral estimate per residue. No boundary padding was applied; consequently, the first and last $\lfloor M/2 \rfloor$ residues of each chain lack STFT values and were excluded from all analyses. For the smallest window ($\sigma = 1.5$, $M = 9$), this trims 4 residues from each end, leaving $\geq 96\%$ of residues in chains of length $\geq 40$. The local power spectral density at residue $n$ is $P(n,\omega_k;\sigma) = |S(n,\omega_k)|^2$, where $S$ denotes the STFT coefficient. The normalized spectral distribution is
	\begin{equation}
		\label{eq:phat}
		\hat{P}(n,\omega_k;\sigma) = \frac{P(n,\omega_k;\sigma)}{\sum_{k'} P(n,\omega_{k'};\sigma)},
	\end{equation}
	and the local spectral entropy is
	\begin{equation}
		\label{eq:Hspec}
		H_{\mathrm{spec}}[n;\sigma] = \frac{-\sum_{k} \hat{P}(n,\omega_k;\sigma) \,\ln\hat{P}(n,\omega_k;\sigma)}{\ln K},
	\end{equation}
	where $K$ is the number of frequency bins, so that $H_{\mathrm{spec}} \in [0,1]$. Residues with zero total power were assigned $H_{\mathrm{spec}} = \mathrm{NaN}$ and excluded from all analyses. To map how the observation scale governs helix--coil separation, the window width was swept over $\sigma \in \{1.5, 2.0, 2.5, 3.0, 3.5, 4.0, 5.0, 6.0, 7.0, 8.0\}$ residues, spanning from the minimum that yields a meaningful spectrum ($M = 9$ frequency bins at $\sigma = 1.5$) to widths exceeding two full helical turns ($\sigma = 8.0$, $M = 48$). For each $\sigma$, the separability between helical and non-helical structural states was quantified by the area under the receiver operating characteristic curve (ROC AUC), which provides a non-parametric measure of the overlap between the probability density distributions of the two structural phases, computed across all aligned residues in the main dataset. Throughout, the ground-truth structural labels for every ROC and AUC calculation are the three-state DSSP assignments \cite{kabsch1983dictionary}, with the helix states (H, G, I) as the positive class and all remaining states as the negative class; the geometric observables enter purely as continuous scores evaluated against these fixed labels. The optimal $\sigma^{*}$ was identified as the value maximizing this phase-state separability, and a cubic spline interpolation over the discrete $\sigma$ grid was used to refine the estimate. To probe the complementary low-pass regime, we defined
	\begin{equation}
		\label{eq:RLF}
		R_{\mathrm{LF}}[n;\sigma,f_c] = \frac{\sum_{|\omega_k| \leq f_c} P(n,\omega_k;\sigma)}{\sum_{k} P(n,\omega_k;\sigma)},
	\end{equation}
	measuring the fraction of local STFT power at or below a frequency cutoff $f_c$. When $f_c = 0$, $R_{\mathrm{LF}}$ reduces to the DC power fraction, which measures how much of the local signal energy is concentrated in the zero-frequency (mean-value) component. In helical segments, where $V_{\mathrm{re}}[n]$ is nearly constant, the DC component dominates and $R_{\mathrm{LF}}$ is large; in coil regions, where $V_{\mathrm{re}}[n]$ fluctuates rapidly, power is distributed across many frequency bins and $R_{\mathrm{LF}}$ is small. The optimal parameters were determined by a systematic parameter sweep over $\sigma \in \{1.5, 2.0, 2.5, \ldots, 8.0\}$ and $f_c$ ranging from 0 (DC only) to the Nyquist frequency in 20 equally spaced steps, maximizing the structural-state separability (ROC AUC) on the main dataset. To quantify the complementarity between the local and macroscopic observables, we formed standardized linear combinations $S[n] = \alpha\,\tilde{X}_1 + (1{-}\alpha)\,\tilde{X}_2$, where $\tilde{X} = (X - \mu_X)/\sigma_X$ is the global $z$-score computed over all aligned residues in the main dataset. The standardization ensures that both components contribute on a common scale regardless of their original units and dynamic ranges. The relative weighting $\alpha^{*}$ was determined by sweeping $\alpha \in [0,1]$ in steps of 0.01 and selecting the value that maximizes phase-state separability (ROC AUC). Two multi-scale observables were constructed: $S_{EH} = \alpha^{*}\tilde{E} + (1{-}\alpha^{*})\tilde{H}_{\mathrm{spec}}$ (combining the local integrability residual with spectral entropy) and $S_{ER} = \alpha^{*}\tilde{E} + (1{-}\alpha^{*})\tilde{R}_{\mathrm{LF}}$ (combining $E[n]$ with the macroscopic energy ratio). Statistical significance of separability improvements relative to $E[n]$ alone was assessed by the DeLong test \cite{delong1988comparing}, which provides a nonparametric comparison of two correlated ROC curves by estimating the variance of the AUC difference through placement values. Because residue-level sample sizes exceed $10^5$, reported $p$-values reflect sample size rather than effect magnitude; practical significance is therefore interpreted through effect sizes (Cohen's $d$, rank-biserial correlation, and AUC differences with 95\% confidence intervals) rather than through $p$-value exponents.
	
	Helix--coil boundaries were identified as positions where the three-state DSSP label changes between H and non-H, yielding a total of 21\,107 such boundaries across the 1\,986 chains; of these, 2\,351 were discarded because they fell within 10 residues of a chain terminus, leaving 18\,756 boundaries with complete $\pm 10$-residue windows (an additional 392 boundaries with partial windows were successfully fitted, yielding 19\,148 valid boundaries in total for the spectral entropy analysis). For each boundary, a sigmoid function $f(n) = L/(1 + e^{-k(n-n_0)}) + b$ was fitted to the $H_{\mathrm{spec}}[n]$ profile in a $\pm 10$-residue window centered on the boundary, using nonlinear least squares (SciPy \cite{virtanen2020scipy}). The four parameters are $L$ (excursion amplitude), $k$ (steepness), $n_0$ (midpoint position), and $b$ (baseline offset). The transition width was defined as the 20\%--80\% excursion width $w = 2\ln 4 / |k|$, which measures the number of residues over which the spectral entropy completes the bulk of its jump. A value $w \ll 1$ indicates a discontinuous transition resembling first-order behavior, while $w \gg 1$ indicates a gradual crossover. The excursion amplitude $|L|$ measures the magnitude of the spectral entropy change across the boundary. Boundaries were classified as helix-entry (C$\to$H) or helix-exit (H$\to$C) to test for directional asymmetry, and the Mann--Whitney $U$ test \cite{mann1947test} with rank-biserial correlation as effect size was used to compare transition width distributions between helix-entry and helix-exit boundaries.

	\begin{figure*}[!t]
		\centering
		\includegraphics[width=0.95\textwidth]{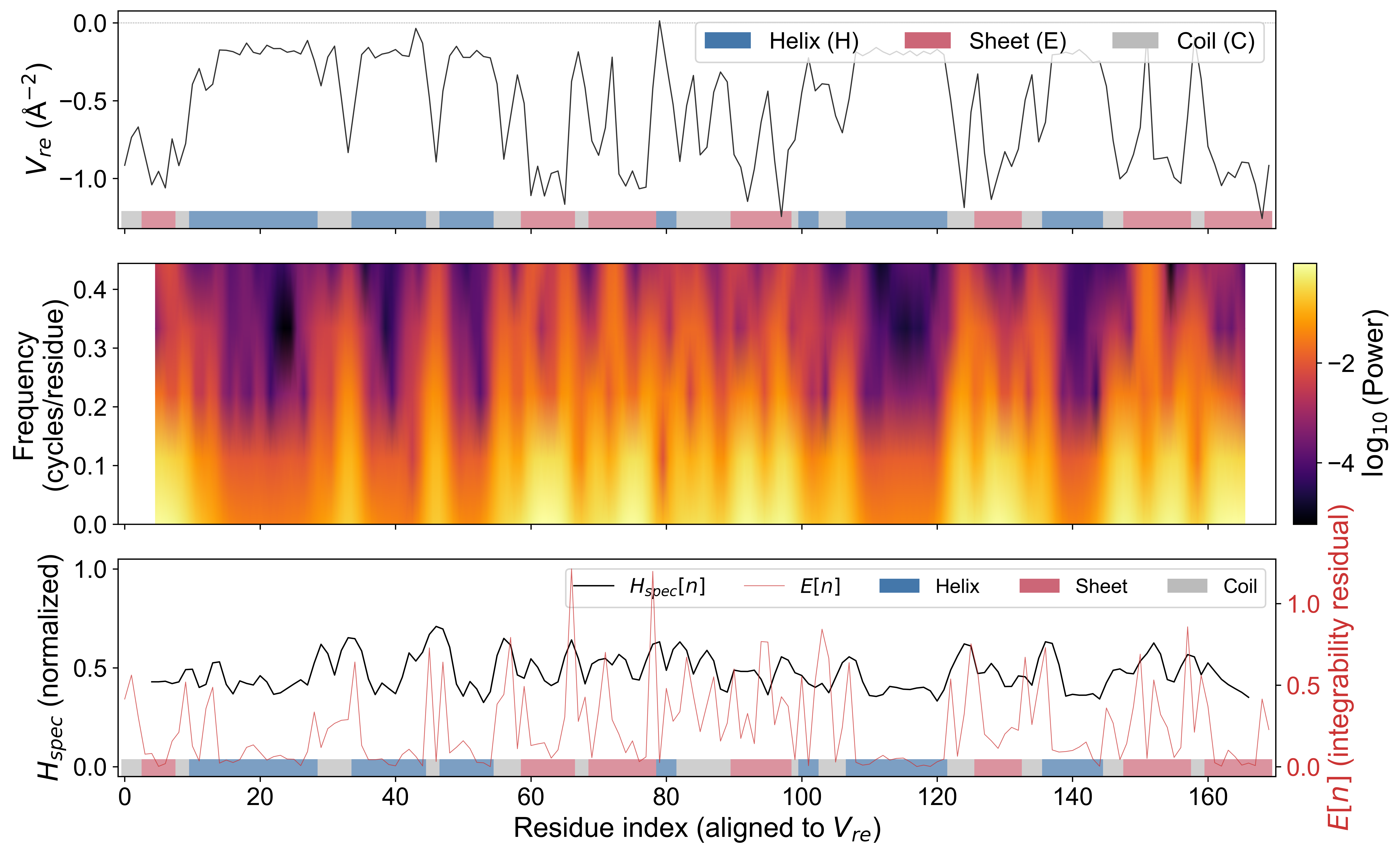}
		\caption{\label{fig:example}Short-time Fourier transform (STFT) analysis of the Hasimoto potential $V_{\mathrm{re}}[n]$ for protein chain 8OSP\_A ($N = 174$ residues, $\sigma = 1.5$). Top panel shows the $V_{\mathrm{re}}[n]$ trace along the backbone ($L = 170$ after end trimming), with DSSP secondary structure annotation shown as a color ribbon (blue = helix, red = sheet, gray = coil). Middle panel shows the STFT power spectrogram $|S(n,\omega)|^2$ computed with a Gaussian window (nperseg $= 9$, 5 frequency bins, 162 time frames), displayed on a $\log_{10}$ scale. Helical regions exhibit concentrated low-frequency power, while coil regions show broadband energy dispersion. Bottom panel shows the normalized spectral entropy $H_{\mathrm{spec}}[n] \in [0,1]$ (black, left axis) and integrability residual $E[n]$ (red, right axis), with 163 of 170 residue positions covered by the STFT window. The mean $H_{\mathrm{spec}}$ follows the ordering helix~(0.444)~$<$~sheet~(0.484)~$<$~coil~(0.523), consistent with the expectation that structurally ordered regions produce lower spectral entropy. The smoother profile of $H_{\mathrm{spec}}$, arising from the STFT window averaging, contrasts with the pointwise sensitivity of $E[n]$; this local and macroscopic complementarity motivates the multi-scale analysis in Sec.~\ref{sec:gabor_limit}.}
	\end{figure*}

	To provide an external reference, these separability values were benchmarked against P-SEA \cite{labesse1997p}, a widely used secondary-structure assigner that operates solely on the C$_\alpha$ trace through fixed distance and angle criteria. P-SEA was applied to the C$_\alpha$ coordinates of each of the 1\,986 chains, and its three-state output was reduced to the same helix-vs-rest partition used throughout, with helix (H, G, I) as the positive class and sheet and coil as the negative class. P-SEA and the geometric observables were scored against identical DSSP labels on the same 320\,453 aligned residues, ensuring a like-for-like comparison. As a member of the C$_\alpha$-only assigner family \cite{khalife2021secondary}, P-SEA provides an independent external reference calibrated to reproduce DSSP-style assignments. On this task P-SEA attains an overall three-state Q3 of 0.749 and a helix-vs-rest accuracy of 0.900. Because P-SEA enforces a minimum helical span of five residues, it does not register short 3$_{10}$ segments, which are nevertheless folded into the helix class under the DSSP reduction; this likely reduces its sensitivity to the shortest helices.
	
	\section{Results}
	\label{sec:results}
	
	\subsection{Order and disorder in backbone geometry}
\label{sec:order_disorder}

Figure~\ref{fig:example} illustrates the STFT analysis for a representative protein chain. The $V_{\mathrm{re}}[n]$ trace shows near-constant negative plateaus in $\alpha$-helical segments and large-amplitude broadband fluctuations in coil regions. The STFT spectrogram $|S(n,\omega;\sigma{=}1.5)|^2$ makes this contrast visible. Helical segments appear as narrow vertical bands concentrated at the DC frequency, while coil regions fill the full frequency axis. The spectral entropy $H_{\mathrm{spec}}[n]$ tracks this distinction as a single scalar, with low values ($H_{\mathrm{spec}} \lesssim 0.3$) within helices rising sharply to $H_{\mathrm{spec}} \gtrsim 0.5$ at helix--coil boundaries.

	\begin{figure*}[!t]
		\centering
		\includegraphics[width=\textwidth]{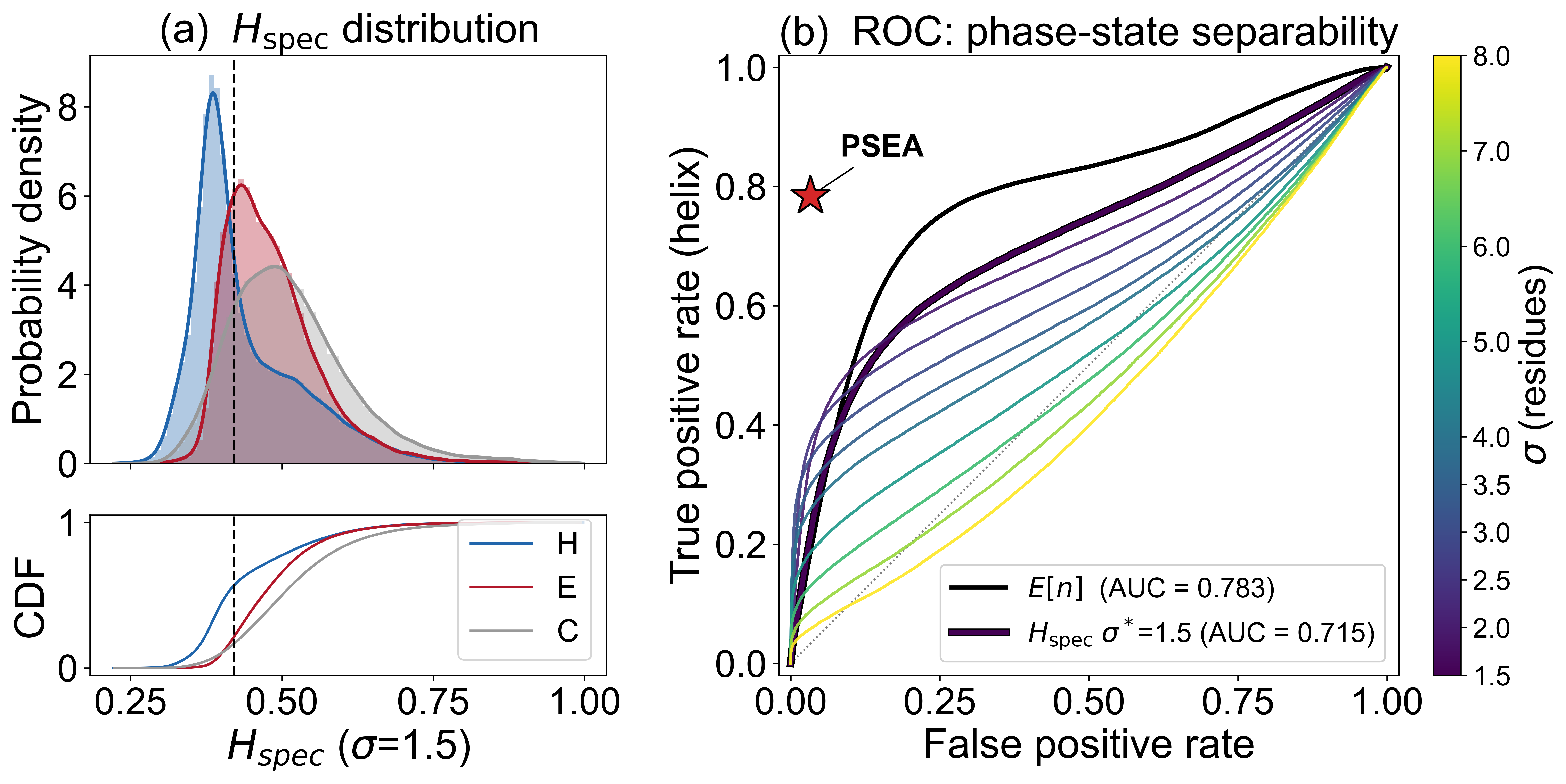}
		\caption{\label{fig:dist_roc}Statistical characterization of spectral entropy $H_{\mathrm{spec}}$ as a structural order parameter across 1\,986 non-redundant protein chains (320\,453 residues). (a)~Left top panel shows the probability density (histogram + KDE) of $H_{\mathrm{spec}}$ at $\sigma = 1.5$ for helix (H, blue; $n = 116\,646$), sheet (E, red; $n = 82\,043$), and coil (C, gray; $n = 121\,764$) residues. The dashed line marks the Youden-optimal threshold $\theta^{*}= 0.421$ \cite{youden1950index}. Cohen's $d$ (H vs.\ non-H) $= 0.652$. Left bottom panel shows the empirical CDF of $H_{\mathrm{spec}}$ by secondary structure class, confirming the stochastic ordering $H \prec E \prec C$. (b)~ROC curves quantifying phase-state separability on the full dataset. The black curve shows the $-E[n]$ baseline (AUC~$= 0.783$); the colored curves show $-H_{\mathrm{spec}}$ at ten window scales spanning $\sigma \in [1.5, 8.0]$, encoded by the color bar. The best spectral-entropy configuration ($\sigma^{*}= 1.5$, AUC~$= 0.715$) is highlighted as a thick black-edged line. The red star marks the single operating point of P-SEA, a C$_\alpha$-only secondary-structure assigner \cite{labesse1997p} evaluated on the identical helix-vs-rest task; as a hard classifier it yields one (FPR,\,TPR) point rather than a curve and has no AUC, reflecting its calibration for DSSP-style assignment. Results on the Wang~(2026) subset are nearly identical (Table~\ref{tab:stratified}, \textit{Origin} row) and are omitted here for clarity. The windowed observable underperforms the pointwise $E[n]$, consistent with the spatial averaging inherent to the STFT; however, this gap is closed by the multi-scale combination (Sec.~\ref{sec:gabor_limit}).}
	\end{figure*}
	
	Across the full dataset of 320\,453 residues, the mean spectral entropy at $\sigma = 1.5$ consistently orders as $\langle H_{\mathrm{spec}}\rangle_{\mathrm{helix}} = 0.440 < \langle H_{\mathrm{spec}}\rangle_{\mathrm{sheet}} = 0.483 < \langle H_{\mathrm{spec}}\rangle_{\mathrm{coil}} = 0.513$ (Fig.~\ref{fig:dist_roc}a). The helix--coil separation has Cohen's $d \approx 0.65$ \cite{cohen2013statistical}, indicating a medium-to-large effect size. This three-level ordering reflects the underlying spectral structure of $V_{\mathrm{re}}[n]$ in each secondary structure class. In helical segments, the near-constant $V_{\mathrm{re}}$ (mean $\mu_H = -0.250$~\AA$^{-1}$, variance $\sigma^2_H = 0.021$~\AA$^{-2}$) concentrates the majority of STFT power at the DC frequency, producing a sharply peaked power spectrum and low entropy. In coil regions, the large-amplitude fluctuations ($\mu_C = -0.611$, $\sigma^2_C = 0.084$) distribute power across the full frequency axis, yielding a flatter spectrum and higher entropy. Sheet residues ($\mu_E = -0.873$, $\sigma^2_E = 0.070$) occupy an intermediate position, with $V_{\mathrm{re}}$ values more negative on average than coil but exhibiting less broadband variability, placing their spectral entropy between the two extremes.
	
	The DC-concentration argument above does not explain why the spectral entropy ordering helix~$<$~sheet~$<$~coil holds in the first place. A naive piecewise-stationary model with white noise predicts an inverted spectral entropy ordering ($H_{\mathrm{spec}}^{\mathrm{helix}} > H_{\mathrm{spec}}^{\mathrm{coil}}$). This discrepancy arises because non-helical segments have a larger mean $|\mu_C|$ in $V_{\mathrm{re}}[n]$ than helical segments ($|\mu_{\mathrm{nonH}}| = 0.716$ vs.\ $|\mu_H| = 0.250$~\AA$^{-1}$). In a white-noise model, this larger DC offset concentrates more power at the zero-frequency bin, which lowers the spectral entropy (making the distribution more peaked) relative to helical segments. Under this model, coil regions would appear more spectrally ordered than helices, contradicting the empirical observation. At $\sigma = 1.5$, the white-noise model predicts $H_{\mathrm{spec}}^{\mathrm{helix}} = 0.598$ and $H_{\mathrm{spec}}^{\mathrm{nonH}} = 0.528$, inverting the empirically observed ranking.
	
	This contradiction is resolved by incorporating the empirical short-range autocorrelation of $V_{\mathrm{re}}[n]$. We fitted an AR(1) colored-noise model with lag-1 autocorrelation $\rho \approx 0.25$ separately to helical and non-helical residues. However, even the AR(1) model fails to reproduce the correct ordering. At $\sigma = 1.5$, it predicts $H_{\mathrm{spec}}^{\mathrm{helix}} = 0.552 > H_{\mathrm{spec}}^{\mathrm{nonH}} = 0.494$, still inverted. The explanation lies instead in the full frequency-domain structure of $V_{\mathrm{re}}[n]$, which these low-order models discard by compressing each class to a mean, a variance, and a single lag-1 correlation. The measured ordering ($H_{\mathrm{spec}}^{\mathrm{helix}} = 0.440 < H_{\mathrm{spec}}^{\mathrm{nonH}} = 0.501$ at $\sigma = 1.5$) is set by the class-specific spectral shape rather than by these summary statistics. The physical origin is that non-helical $V_{\mathrm{re}}$ fluctuations are genuinely broadband (colored noise with significant power across all frequencies), not merely large-amplitude white noise. The empirical spectral flatness of non-helical residues exceeds that of helical residues at every observation scale (0.89 versus 0.72 at $\sigma = 1.5$, and 0.69 versus 0.46 at $\sigma = 8$), confirming that coil regions distribute power more uniformly across frequencies. It is this spectral bandwidth difference, rather than the DC level alone, that drives the entropy separation between secondary structure types, indicating that the ordering is a non-trivial property of the backbone geometry rather than a simple consequence of signal amplitude differences.
	
	The ordering is robust, holding across all 1\,986 individual chains and persisting across all stratification layers (by helix content, chain length, and dataset origin; Table~\ref{tab:stratified}). ROC analysis, taking the three-state DSSP labels as ground truth, confirms that both observables distinguish helical from non-helical structural states. In this context, an AUC of 0.5 corresponds to completely overlapping probability density distributions (indistinguishable structural states), while values approaching 1.0 indicate well-separated distributions and strong symmetry breaking between the ordered and disordered phases. The pointwise integrability residual $E[n]$ achieves a phase-state separability of AUC~$= 0.783$, while the windowed spectral entropy $H_{\mathrm{spec}}$ at $\sigma^{*} = 1.5$ yields AUC~$= 0.715$ (Fig.~\ref{fig:dist_roc}b; Table~\ref{tab:stratified}). The systematically higher separability of $E[n]$ across all strata reflects its pointwise sensitivity to local geometric discontinuities, a consequence of operating without a spatial averaging window.

	These separability values acquire quantitative meaning through comparison with an external reference. At its single operating point the C$_\alpha$-only assigner P-SEA reaches a sensitivity of 0.784 and a specificity of 0.967 against the same DSSP labels (Table~\ref{tab:headtohead}), and this operating point lies above the geometric ROC curves in Fig.~\ref{fig:dist_roc}b. This ordering is expected. P-SEA encodes distance and angle criteria calibrated specifically to reproduce DSSP-style helix assignments, so it is optimized for exactly this discrete classification task, whereas $E[n]$ and $H_{\mathrm{spec}}$ are continuous physical order parameters derived from the effective potential and are never tuned for assignment. Interpreted this way, the AUC values report how much of the discriminative power of a purpose-built assigner is recovered by untuned geometric order parameters, and they are anchored simultaneously by the chance level of 0.5, by the internal $E[n]$ reference, and by the external P-SEA operating point.

	\begin{table}[htbp]
		\caption{\label{tab:headtohead}Head-to-head comparison against the P-SEA C$_\alpha$-only baseline on the helix-versus-rest task (320\,453 residues, DSSP labels). The three geometric scores are thresholded at their Youden-optimal operating points \cite{youden1950index}. All metrics refer to the helix-positive partition.}
		\footnotesize
		\begin{tabular}{lcccc}
			\toprule
			Method & Sensitivity & Specificity & Helix F1 & MCC \\
			\midrule
			P-SEA \cite{labesse1997p} & 0.784 & 0.967 & 0.851 & 0.784 \\
			$E[n]$ & 0.722 & 0.783 & 0.688 & 0.497 \\
			$H_{\mathrm{spec}}$ & 0.567 & 0.814 & 0.600 & 0.392 \\
			$S_{EH}$ & 0.665 & 0.848 & 0.689 & 0.521 \\
			\bottomrule
		\end{tabular}
	\end{table}
	
	\begin{table}[htb]
		\caption{\label{tab:stratified}Stratified phase-state separability (ROC AUC) for $E[n]$ and $H_{\mathrm{spec}}[n;\sigma{=}1.5]$ across dataset subgroups. $\Delta$AUC $= \mathrm{AUC}_{H} - \mathrm{AUC}_{E}$ is negative across all strata, reflecting the superior boundary sensitivity of the pointwise operator.}
		\begin{tabular}{llccc}
				\toprule
				Stratum & Group & AUC($E$) & AUC($H_{\mathrm{spec}}$) & $\Delta$AUC \\
				\midrule
				Overall & All & 0.783 & 0.715 & $-$0.068 \\
				\midrule
				Helix \% & Low & 0.664 & 0.629 & $-$0.035 \\
				& Mid & 0.772 & 0.706 & $-$0.066 \\
				& High & 0.823 & 0.774 & $-$0.049 \\
				\midrule
				Length & Short & 0.813 & 0.756 & $-$0.057 \\
				& Medium & 0.782 & 0.717 & $-$0.065 \\
				& Long & 0.779 & 0.706 & $-$0.073 \\
				\midrule
				Origin & Ref.~\cite{wang2026structural} & 0.782 & 0.711 & $-$0.071 \\
				& New & 0.784 & 0.718 & $-$0.066 \\
				\bottomrule
			\end{tabular}
	\end{table}
	
	\subsection{Single-step boundary sharpness and thermodynamic cooperativity}
\label{sec:boundary_sharpness}

	\begin{figure*}[!t]
		\centering
		\includegraphics[width=0.9\textwidth]{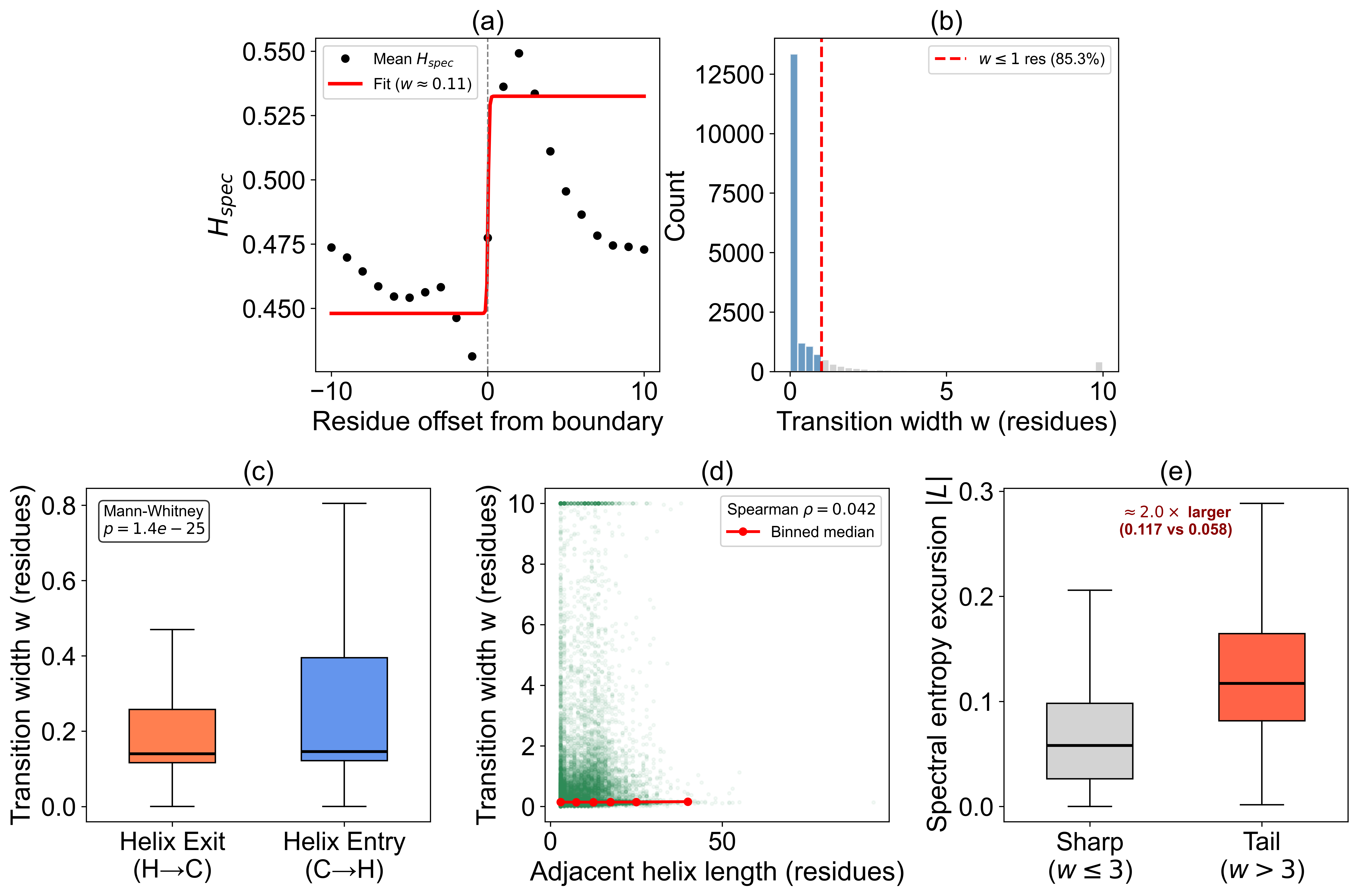}
		\caption{Abrupt geometric transition at helix--coil boundaries. (a) Representative sigmoid fit (red curve) to the local spectral entropy $H_{\mathrm{spec}}$ profile (black dots) at a single helix--coil boundary. (b) Distribution of the fitted transition widths $w$ across 19\,148 boundaries. The vertical dashed line marks $w = 1$ residue; 85.3\% of boundaries fall below this single-residue limit. (c) Directional asymmetry in transition sharpness. Helix-exit (H$\to$C) boundaries are sharper on average than helix-entry (C$\to$H) transitions (Mann--Whitney test, rank-biserial $r = 0.087$). (d) Transition width $w$ shows no meaningful correlation with the adjacent helix length (Spearman $\rho = 0.042$). (e) Physical characterization of the distribution tail ($w > 3$ residues). These gradual boundaries exhibit a spectral entropy excursion $|L|$ approximately 2.0 times larger (median $\approx 0.117$) than that of the sharp majority (median $\approx 0.058$), indicating that the broader spatial transition is associated with a larger absolute jump in local spectral disorder.}
		\label{fig:phase}
	\end{figure*}

The distinct spatial resolutions of the local and macroscopic observables motivate a direct characterization of the helix--coil phase-boundary width. Sigmoid fits to the 19,148 valid helix--coil boundaries across the 1,986 chains reveal that the spectral entropy transition is sharp (Fig.~\ref{fig:phase}a). The median transition width is $w = 0.145$ residues, and 85.3\% of boundaries have $w \leq 1$ residue (Fig.~\ref{fig:phase}b). A further 8.5\% fall in the range $1 < w \leq 3$ residues, and only 6.2\% exceed 3 residues. Because $V_{\mathrm{re}}[n]$ is sampled at integer C$_\alpha$ positions, the physically resolvable transition width is bounded below by one residue spacing; the value $w = 0.145$ is a parameter of the continuous sigmoid fitted to these discrete samples and quantifies the steepness of a single-step transition, meaning that the spectral entropy jump is effectively complete between one residue and the next, rather than a literal sub-residue spatial extent. This single-step sharpness is characteristic of an abrupt, first-order-like transition rather than a gradual crossover, providing an independent kinematic counterpart to the high thermodynamic cooperativity (small nucleation parameter $\sigma_{\mathrm{ZB}} \ll 1$) of the Zimm--Bragg helix--coil model \cite{zimm1959theory}. The median sigmoid steepness parameter is $k = 19.1$.

	A directional asymmetry is observed, with helix-exit (H$\to$C) boundaries being sharper than helix-entry (C$\to$H) boundaries (median transition widths 0.142 and 0.149 residues respectively; rank-biserial $r = 0.087$, Mann--Whitney test). Although the effect size is small in absolute terms, the direction is consistent and has a natural structural basis in helix capping, where both termini carry distinctive local structure rather than fraying gradually into disorder \cite{aurora1998helix, segura2012caps}. The two ends are capped through structurally different means. Carboxyl termini are frequently closed by glycine-rich Schellman and $\alpha_L$ motifs in which the terminating residue adopts a positive-$\phi$ backbone conformation that lies well outside the helical region of the Ramachandran map \cite{schellman1980alphal, viguera1995experimental, serrano1992alpha, sakuma2024design}. Because this $\alpha_L$ state is a discrete non-helical conformation rather than a flexibility-driven continuous relaxation, the backbone trace turns abruptly at the carboxyl terminus rather than relaxing smoothly \cite{kallenbach1999c, newell2015mapping}. Amino termini, by contrast, are stabilized largely by side-chain to backbone hydrogen bonds from short polar residues such as asparagine, serine, and threonine \cite{doig1997structures, vijayakumar1999hydrogen}, a sequence-level feature that need not perturb the backbone trace as sharply \cite{penel1999side, doig1995n, kargatov2024strained}. A slightly sharper exit than entry is therefore the static geometric signature consistent with this difference in terminal architecture. The continuous, sequence-agnostic order parameter used here registers this small asymmetry directly from the C$_\alpha$ trace, complementing rather than duplicating the sequence and hydrogen-bond descriptions on which the capping literature is built and connecting to established continuous characterizations of helix geometry \cite{hauser2017characterization, bansal2000helanal, kumar2012helanal}.

	The 6.2\% of boundaries with $w > 3$ residues (the distribution tail) do not correspond to longer helices or higher integrability residuals, as their median helix length and mean $E[n]$ are indistinguishable from the sharp majority. However, their spectral entropy excursion $|L|$ is approximately 2.0 times larger than that of sharp boundaries (0.117 vs.\ 0.058), indicating that these gradual transitions involve a larger absolute change in local spectral character spread over more residues. Among tail boundaries, C$\to$H entries are overrepresented (59\%), consistent with the directional asymmetry noted above.

	The sharpness is effectively invariant to helix type ($\alpha$, $3_{10}$, $\pi$), adjacent secondary structure element, and helix length. Although Kruskal--Wallis tests \cite{kruskal1952use} reach nominal significance for helix type and helix length, the effect sizes are negligible (rank-biserial $|r| < 0.02$; Spearman $\rho = 0.042$), suggesting that sharpness is largely an intrinsic geometric property of the helix--coil interface rather than a context-dependent feature.

	\subsection{Resolution regimes of pointwise and windowed spectral probes}
\label{sec:gabor_limit}

The single-step sharpness of these phase boundaries reveals a resolution constraint inherent to any windowed spectral method.

	\begin{figure*}[!t]
		\centering
		\includegraphics[width=0.9\textwidth]{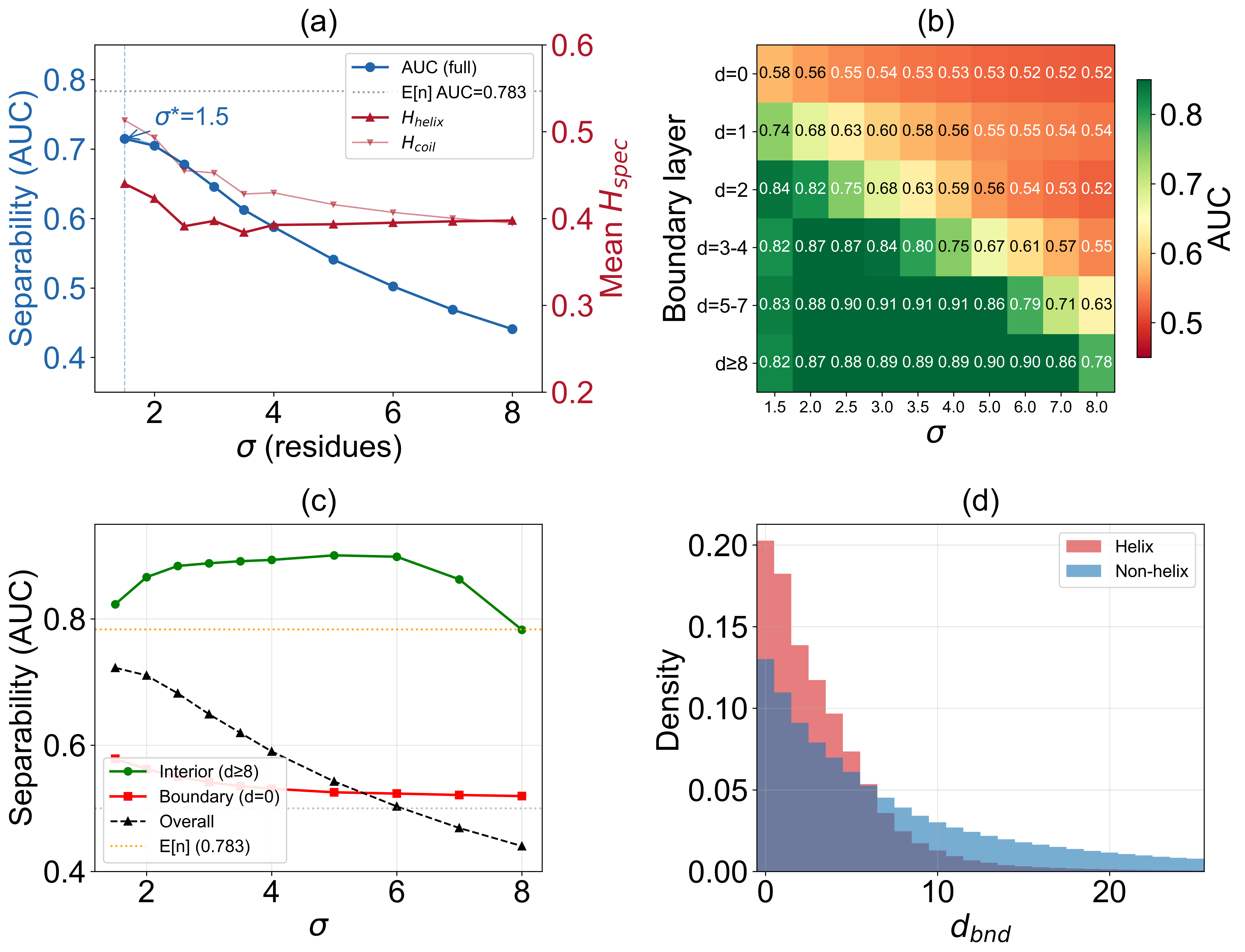}
		\caption{\label{fig:sigma_bnd}Scale-dependent boundary contamination and the loss of spatial resolution. (a)~Phase-state separability (ROC AUC, blue, left axis) and mean spectral entropy $H_{\mathrm{spec}}$ (red, right axis) as functions of the observation scale $\sigma$. The full-residue AUC declines monotonically with increasing $\sigma$, while the mean $H_{\mathrm{spec}}$ for helix and coil residues converges. The grey dotted line marks the pointwise $E[n]$ baseline (AUC $= 0.783$). The optimal observation scale is $\sigma^* \approx 1.5$ residues (AUC $= 0.715$). (b)~Heatmap of phase-state separability across boundary layers (rows) and $\sigma$ values (columns). High separability (green) is concentrated in interior layers at small $\sigma$; resolution degrades toward the phase boundary and at larger observation scales. (c)~Comparison of separability for interior ($d_{\mathrm{bnd}} \geq 8$, green), boundary ($d_{\mathrm{bnd}} = 0$, red), and overall (black dashed) residues as a function of $\sigma$. Interior separability peaks near 0.82 at small $\sigma$, while boundary separability remains near 0.52--0.58, confirming that phase-boundary contamination is the dominant source of resolution loss. The orange dotted line indicates the pointwise $E[n]$ baseline (AUC $= 0.783$). (d)~Distribution of boundary distance $d_{\mathrm{bnd}}$ for helix (red) and non-helix (blue) residues. Both distributions are right-skewed, with a substantial fraction of residues located near helix--coil phase boundaries ($d_{\mathrm{bnd}} \leq 4$), where the windowed spectral probe suffers resolution loss from boundary contamination.}
	\end{figure*}
	
	The dependence of structural resolving power on the observation scale $\sigma$ is shown in Figure~\ref{fig:sigma_bnd}a. The AUC decreases monotonically from 0.715 at $\sigma = 1.5$ to 0.441 at $\sigma = 8.0$, with no local maximum at any intermediate scale. The mean $H_{\mathrm{spec}}$ for helix and coil residues converges with increasing $\sigma$, reflecting the progressive loss of spectral contrast as the window averages over increasingly heterogeneous geometry. At $\sigma \geq 3.0$, the Gaussian window extends beyond the ends of short chains, producing undefined values for edge residues. The NaN fraction grows from 10.5\% at $\sigma = 3.0$ to 29.0\% at $\sigma = 8.0$, further degrading the practical utility of wide windows. The absence of a peak near one helical period ($\sigma \approx 3.6$) indicates that the helix--coil discrimination does not rely on resolving the periodic structure of helices; rather, it depends on detecting the local spectral contrast between ordered and disordered geometry.
	
	This monotonic decay can be understood through a piecewise-stationary signal model. We model $V_{\mathrm{re}}[n]$ as alternating segments of constant mean $\mu_H$ (helix) and $\mu_C$ (coil) with additive noise. The intrinsic discriminability between the two classes, measured by the squared standardized mean difference $d^2_{\mathrm{int}}(\sigma)$, increases with $\sigma$ because wider windows average out noise. However, the fraction of residues whose STFT window straddles a helix--coil boundary also grows. At $\sigma = 1.5$, approximately 53\% of residues are boundary-contaminated (within $3\sigma$ of a transition), rising to 88\% at $\sigma = 8.0$. The effective discriminant $d^2_{\mathrm{eff}} \propto (1 - f_{\mathrm{bnd}})^2 \, d^2_{\mathrm{int}}$ is dominated by the contamination fraction and decreases monotonically, reproducing the observed AUC trend (Spearman $\rho = 0.81$, $p = 0.004$).
	
	Stratified validation by boundary distance $d_{\mathrm{bnd}}$ confirms this picture (Fig.~\ref{fig:sigma_bnd}b,\,c). For boundary residues ($d_{\mathrm{bnd}} = 0$), AUC is near chance (0.52--0.58) at all $\sigma$. For interior residues ($d_{\mathrm{bnd}} = 5$--7), AUC peaks at $\sigma = 3.5$ (AUC~$= 0.912$), while for deeply buried residues ($d_{\mathrm{bnd}} \geq 8$) the peak shifts to $\sigma = 5.0$ (AUC~$= 0.901$). The rightward shift of the optimal $\sigma$ with increasing $d_{\mathrm{bnd}}$ reflects the balance between frequency-resolution gain and boundary contamination, as deeper interior residues tolerate wider windows before the nearest boundary enters the $3\sigma$ support. This imposes a maximal-localization constraint in which the overall AUC is maximized at the smallest $\sigma$ because the majority of residues in real proteins lie close to a helix--coil boundary.
	
	To quantify this limit, we model the helix--coil boundary as an ideal step in $V_{\mathrm{re}}[n]$, with $V[n] = V_0 + \eta_H[n]$ for $n < n_0$ (helical plateau) and $V[n] = \mu_C + \eta_C[n]$ for $n \geq n_0$ (coil mean), where $\eta_{H/C}[n]$ are zero-mean local fluctuations with variances $\sigma_{H/C}^2$. When the STFT window of width $\sigma$ is centered at position $\xi = (n_0 - n_{\mathrm{center}})/\sigma$ from the boundary, the expected power spectrum of the windowed signal takes the form
	\begin{multline}
	\label{eq:Pk_boundary}
	\mathbb{E}[P_k(\xi)] = \bigl|\Delta V \cdot G_H(\omega_k;\xi) + \mu_C\, G(\omega_k)\bigr|^2 \\
	+ \alpha(\xi)\,\sigma_H^2\, W_E\, S_k^{(H)} + [1{-}\alpha(\xi)]\,\sigma_C^2\, W_E\, S_k^{(C)},
	\end{multline}
	where $\Delta V = V_0 - \mu_C = 0.361$~\AA$^{-1}$ is the mean step, $G_H(\omega_k;\xi)$ is the partial-window DFT over the helical portion, $G(\omega_k)$ is the full-window DFT, $\alpha(\xi)$ is the fraction of window energy in the helical region, $W_E$ is the total window energy, and $\{S_k^{(H,C)}\}$ are the measured fluctuation spectral shapes. The first term captures spectral leakage from the mean discontinuity. Because the half-Gaussian window $G_H$ lacks the frequency concentration of the full window $G$, its Fourier coefficients decay as $\sim 1/k$ rather than exponentially, redistributing energy from the DC bin into higher frequencies. This leakage produces a characteristic entropy peak at the boundary (Fig.~S1a).

	Numerical evaluation of Eq.~(\ref{eq:Pk_boundary}) with the empirical parameters ($V_0 = -0.250$, $\mu_C = -0.611$, $\sigma_H^2 = 0.021$, $\sigma_C^2 = 0.084$, $\sigma = 1.5$, $K = 5$) predicts a boundary entropy peak of $H_{\mathrm{spec}}^{\mathrm{peak}} = 0.572$ at a position 1.0~residue into the coil side, with a full-width at half-maximum (FWHM) of 2.0~residues. These predictions are confirmed by the empirical mean boundary profile averaged over 500 helix--coil transitions: peak $= 0.563$ at position $+1$, FWHM $= 2$~residues (Fig.~S1b). The theoretical FWHM exceeds the measured geometric boundary width ($w = 0.145$~residues) by a factor of nearly 14, indicating that the single-residue-step helix--coil discontinuity cannot be resolved by any finite-window spectral method. This resolution gap is consistent with the Gabor uncertainty principle \cite{gabor1946electrical}. Meaningful spectral discrimination requires $\sigma \geq 1$~residue (to produce $K \geq 4$ frequency bins), which imposes a minimum transition width of $\mathrm{FWHM} \sim O(\sigma)$. The pointwise integrability residual $E[n]$, operating without an integration window, is not subject to this constraint and achieves the lattice-limited resolution of $\sim 0.15$~residues. This resolution asymmetry provides the theoretical foundation for the multi-scale framework. The observables $H_{\mathrm{spec}}$ and $E[n]$ are not merely empirically complementary but occupy distinct resolution regimes, with the windowed $H_{\mathrm{spec}}$ bound by the Gabor window trade-off and the windowless $E[n]$ operating at the lattice limit.

	\begin{figure*}[!t]
		\centering
		\includegraphics[width=\textwidth]{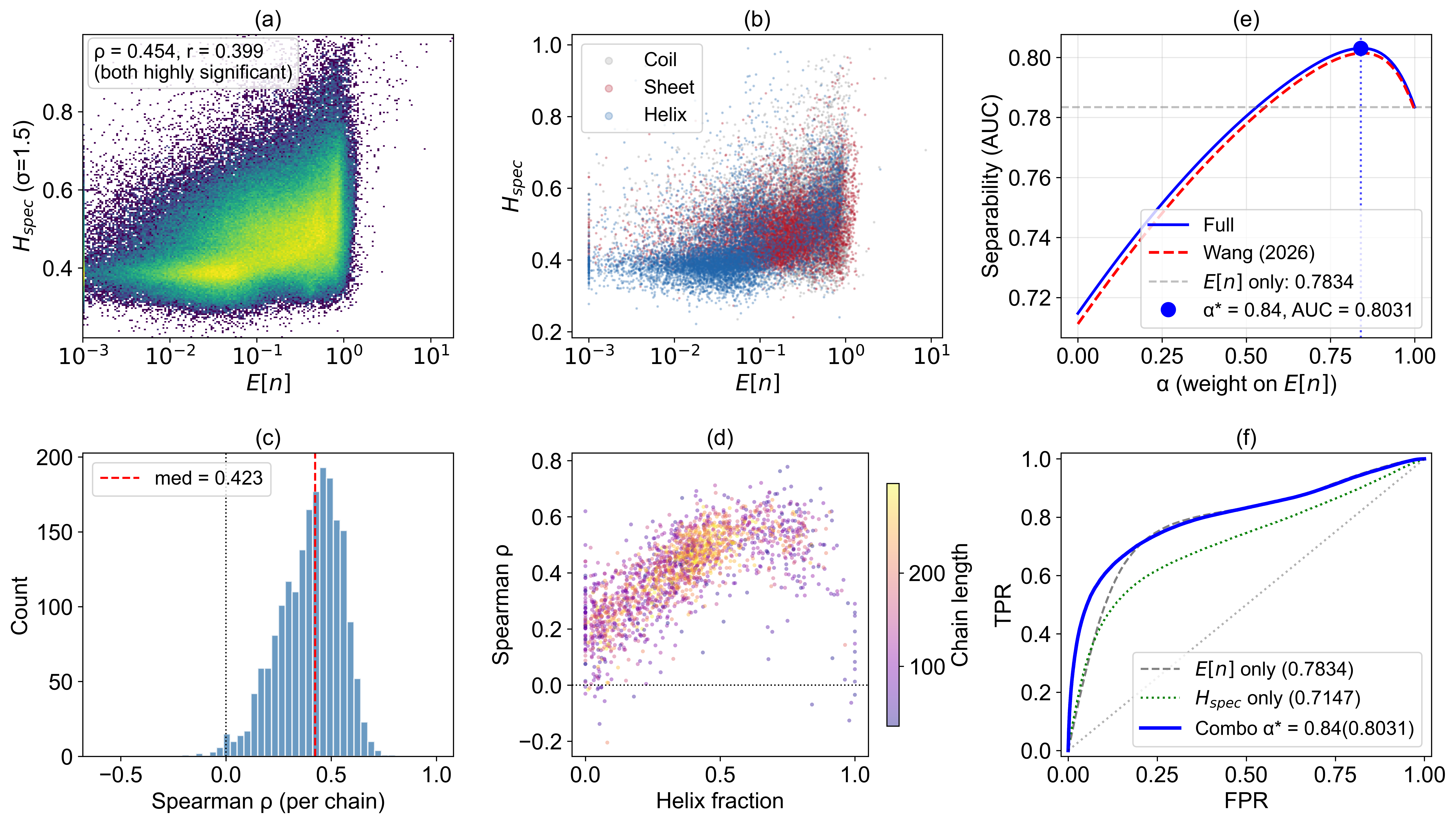}
		\caption{Correlation between spectral entropy $H_{\mathrm{spec}}$ and integrability residual $E[n]$, and their combined structural resolving power. (a)~Density scatter plot of $H_{\mathrm{spec}}$ ($\sigma^* = 1.5$) versus $E[n]$ for all 320,453 valid residues (log-scaled x-axis). Global Spearman $\rho = 0.454$ and Pearson $r = 0.399$ (both highly significant). (b)~Same scatter colored by DSSP secondary-structure assignment, with helix (H, blue), sheet (E, red), and coil (C, gray); per-class Spearman $\rho = 0.472$ (H), $0.261$ (E), $0.147$ (C). (c)~Histogram of per-chain Spearman $\rho$ across 1,986 chains (median $= 0.423$; 99.1\% of chains show $\rho > 0$; 76.2\% show $\rho > 0.3$). (d)~Per-chain Spearman $\rho$ versus helix fraction, colored by chain length; positive correlation is strongest for helix-rich chains. (e)~Phase-state separability (ROC AUC) as a function of the relative weighting $\alpha$ in the combined multi-scale observable $S^* = \alpha\, \tilde{E} + (1-\alpha)\, \tilde{H}$. The optimum $\alpha^* = 0.84$ yields AUC $= 0.803$ on the full dataset (blue solid) and $\alpha^* = 0.85$, AUC $= 0.802$ on the Wang (2026) subset (red dashed), compared with AUC $= 0.783$ for $E[n]$ alone (gray dashed). (f)~ROC curves at the optimal weighting, where the combined observable (blue solid, AUC $= 0.803$) outperforms both the local $E[n]$ limit (gray dashed, AUC $= 0.783$) and the macroscopic $H_{\mathrm{spec}}$ limit (green dotted, AUC $= 0.715$).}
		\label{fig:corr_combo}
	\end{figure*}
	
	The residue-level Spearman rank correlation between $H_{\mathrm{spec}}[n;\sigma{=}1.5]$ and $E[n]$ across all 320\,453 residues is $\rho = 0.454$ (highly significant; Fig.~\ref{fig:corr_combo}a). This moderate value indicates that the two indicators capture related but non-redundant geometric information. The relationship is expected because both quantities respond to the spectral structure of $V_{\mathrm{re}}[n]$. The operator $E[n]$ is a pointwise integrability probe sensitive exclusively to nearest-neighbor torsion changes (under uniform curvature, $E[n] \approx \tfrac{1}{2}|\cos\tau[n] - \cos\tau[n{+}1]|$), whereas $H_{\mathrm{spec}}$ integrates spectral content over a $6\sigma \approx 9$-residue window and summarizes the full bandwidth of the local power spectrum.
	
	Per-chain analysis on the 251 helical peptides yields a mean Spearman $\rho = 0.424$ (median 0.487), with 94.8\% of chains showing $\rho > 0$, confirming that the positive association is consistent across individual proteins. The nature of the non-redundant information is clarified by a segment-level analysis using the Jensen--Shannon divergence (JSD) \cite{lin2002divergence} between the STFT power spectrum of each helical segment and the ideal DNLS soliton spectrum. The JSD correlates strongly with $H_{\mathrm{spec}}$ ($\rho = 0.800$) but only moderately with $E[n]$ ($\rho = 0.522$), and partial correlation analysis confirms that JSD carries independent frequency-domain information beyond $E[n]$. The partial correlation of JSD vs.\ $H_{\mathrm{spec}}$ controlling for $E[n]$ is $+0.766$, whereas JSD vs.\ $E[n]$ controlling for $H_{\mathrm{spec}}$ drops to $+0.407$. The JSD serves as a close proxy for spectral flatness ($\rho = 0.948$), indicating that the non-redundant content of $H_{\mathrm{spec}}$ relative to $E[n]$ is predominantly the broadband character of the local power spectrum rather than any single-frequency feature.
	
	Given this theoretical resolution asymmetry and this moderate correlation, we quantify the complementarity by forming a standardized linear combination $S = \alpha\,\tilde{E} + (1{-}\alpha)\,\tilde{H}$, where $\tilde{E}$ and $\tilde{H}$ denote globally standardized (zero mean, unit variance) versions of $-E[n]$ and $-H_{\mathrm{spec}}[n]$ respectively (negated so that higher values indicate helix). The relative weighting was optimized over $\alpha \in [0,1]$ in steps of 0.01, yielding $\alpha^{*} = 0.84$ and a phase-state separability of AUC~$= 0.803$ ($\Delta\mathrm{AUC} = +0.020$ over $E[n]$ alone, DeLong 95\% CI $[+0.019, +0.020]$; Fig.~\ref{fig:corr_combo}e,\,f). The high optimal weight on $E[n]$ ($\alpha^{*} = 0.84$) reflects the dominance of the local order parameter. Although the absolute gain is modest ($\Delta\mathrm{AUC} = +0.020$), it is consistent across all six stratification layers ($\Delta\mathrm{AUC}$ ranges from $+0.015$ to $+0.026$; Table~\ref{tab:combo}) and confirms that the macroscopic spectral probe contributes non-redundant structural information.
	
	\begin{table}[htb]
		\caption{\label{tab:combo}Phase-state separability across diverse structural ensembles. $S_{EH} = 0.84\,\tilde{E} + 0.16\,\tilde{H}$ and $S_{ER} = 0.81\,\tilde{E} + 0.19\,\tilde{R}_{\mathrm{LF}}$. $\Delta$AUC (separability gain) is relative to $E[n]$ alone in each stratum. Strata are defined by terciles of chain length and helix fraction.}
		\footnotesize
		\begin{tabular}{lccc}
				\toprule
				Stratum & AUC($E$) & $S_{EH}$ / $\Delta$ & $S_{ER}$ / $\Delta$ \\
				\midrule
				Overall & 0.783 & 0.803 / +0.020 & 0.815 / +0.032 \\
				Helix\% Low & 0.696 & 0.711 / +0.015 & 0.728 / +0.032 \\
				Helix\% High & 0.810 & 0.836 / +0.026 & 0.846 / +0.036 \\
				Short chains & 0.797 & 0.820 / +0.023 & 0.828 / +0.031 \\
				Long chains & 0.779 & 0.798 / +0.019 & 0.811 / +0.032 \\
				Ref.~\cite{wang2026structural} subset & 0.782 & 0.801 / +0.019 & 0.814 / +0.032 \\
				\bottomrule
			\end{tabular}
	\end{table}
	
	The connection to soliton physics deepens the interpretation of this complementarity. As shown above, the JSD between each helical segment spectrum and the ideal soliton reference tracks the Wiener spectral flatness \cite{wiener1949extrapolation} and carries frequency-domain information beyond $E[n]$; controlling for JSD reduces the $E[n]$--$H_{\mathrm{spec}}$ correlation to $-0.095$, confirming that their shared content is predominantly the broadband spectral character.
	
	\begin{figure*}[!t]
		\centering
		\includegraphics[width=\textwidth]{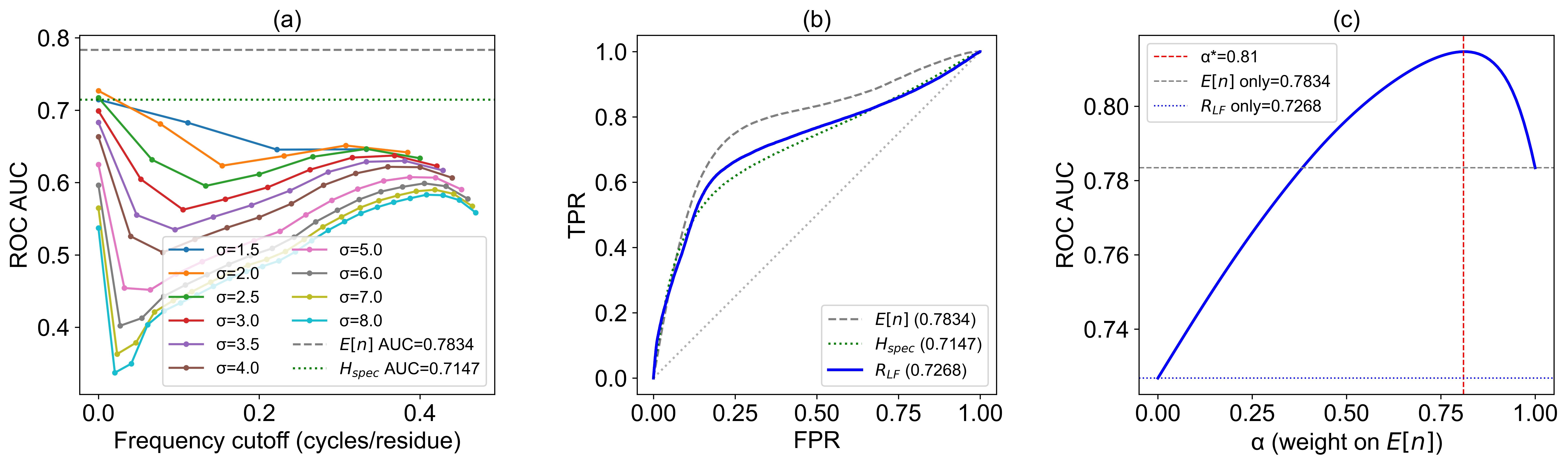}
		\caption{Low-frequency energy ratio $R_{\mathrm{LF}}$ as an indicator of helical order. (a) Phase-state separability (ROC AUC) of $R_{\mathrm{LF}}$ as a function of frequency cutoff $f_c$ for ten observation scales ($\sigma = 1.5$--$8.0$). Dashed gray and dotted green lines mark the reference separability limits of $E[n]$ (0.7834) and $H_{\mathrm{spec}}$ (0.7147), respectively. For all $\sigma$, the best separability is achieved at $f_c = 0$ (DC-only bin), with $\sigma = 2.0$ yielding the highest single-cutoff AUC of 0.7268. Larger $\sigma$ values progressively degrade resolution, consistent with over-smoothing of the local spectral structure. (b) ROC curves for the best $R_{\mathrm{LF}}$ configuration ($\sigma = 2.0$, $f_c = 0.000$) compared with $E[n]$ and $H_{\mathrm{spec}}$ reference limits. $R_{\mathrm{LF}}$ alone (AUC $= 0.7268$) falls below $E[n]$ (0.7834) but slightly exceeds $H_{\mathrm{spec}}$ (0.7147). (c) Separability of the combined observable $S = \alpha \cdot \tilde{E} + (1-\alpha) \cdot \tilde{R}_{\mathrm{LF}}$ as a function of the relative weighting $\alpha$. The optimum at $\alpha^* = 0.81$ achieves AUC $= 0.8147$, a gain of $+0.0313$ over $E[n]$ alone, suggesting that $R_{\mathrm{LF}}$ captures complementary spectral information not fully encoded in the pointwise local probe.}
		\label{fig:lowfreq}
	\end{figure*}
	
	Optimizing $R_{\mathrm{LF}}[n;\sigma,f_c]$ (Eq.~\ref{eq:RLF}) over $\sigma$ and $f_c$ for phase-state separability yields $\sigma = 2.0$ and $f_c = 0$ (DC bin only), with a standalone AUC~$= 0.727$. This is below $E[n]$ (0.783) but above $H_{\mathrm{spec}}$ (0.715), indicating that the DC fraction alone carries substantial structural information. The multi-scale combination $S_{ER} = 0.81\,\tilde{E} + 0.19\,\tilde{R}_{\mathrm{LF}}$ achieves AUC~$= 0.815$ ($\Delta\mathrm{AUC} = +0.031$, DeLong 95\% CI $[+0.031, +0.032]$; Fig.~\ref{fig:lowfreq}), surpassing the $E$--$H_{\mathrm{spec}}$ combination. The optimality of $f_c = 0$ indicates that the structural information in the low-frequency regime is concentrated at zero frequency, because helical segments exhibit a dominant DC component when their $V_{\mathrm{re}}[n]$ is governed by the near-constant integrable plateau. The local order parameter $E[n]$ and macroscopic spectral probe $R_{\mathrm{LF}}$ thus form a complementary pair of multi-scale structural observables.
	
	The local order parameter $E[n]$ and the windowed spectral entropy $H_{\mathrm{spec}}$ capture substantially complementary structural information at different spatial scales. To illustrate the 3D consequences of this complementarity, we reconstruct 3D backbone geometry from the segmented signals \cite{wang2026piecewise} and use the 3D root-mean-square deviation (RMSD) as a macroscopic readout that directly reflects the resolution trade-off between the pointwise and windowed spectral probes.

	Applying the $E[n] > 0.10$ threshold on the local order parameter yields a global topological reconstruction success rate of 72.9\% (RMSD~$< 2$~\AA) on the 251 helical peptide dataset. In contrast, the optimally thresholded macroscopic spectral indicator ($H_{\mathrm{spec}} > \theta^{*}$) achieves only 57.4\%. This 15.5 percentage-point gap illustrates the macroscopic 3D consequence of the resolution limit of the windowed probe. The STFT window blurs the single-residue-step phase boundaries that $E[n]$ resolves with lattice-level fidelity, leading to terminal fraying and degraded RMSD in globally clean helices.
	
	\begin{figure*}[!t]
		\centering
		\includegraphics[width=\textwidth]{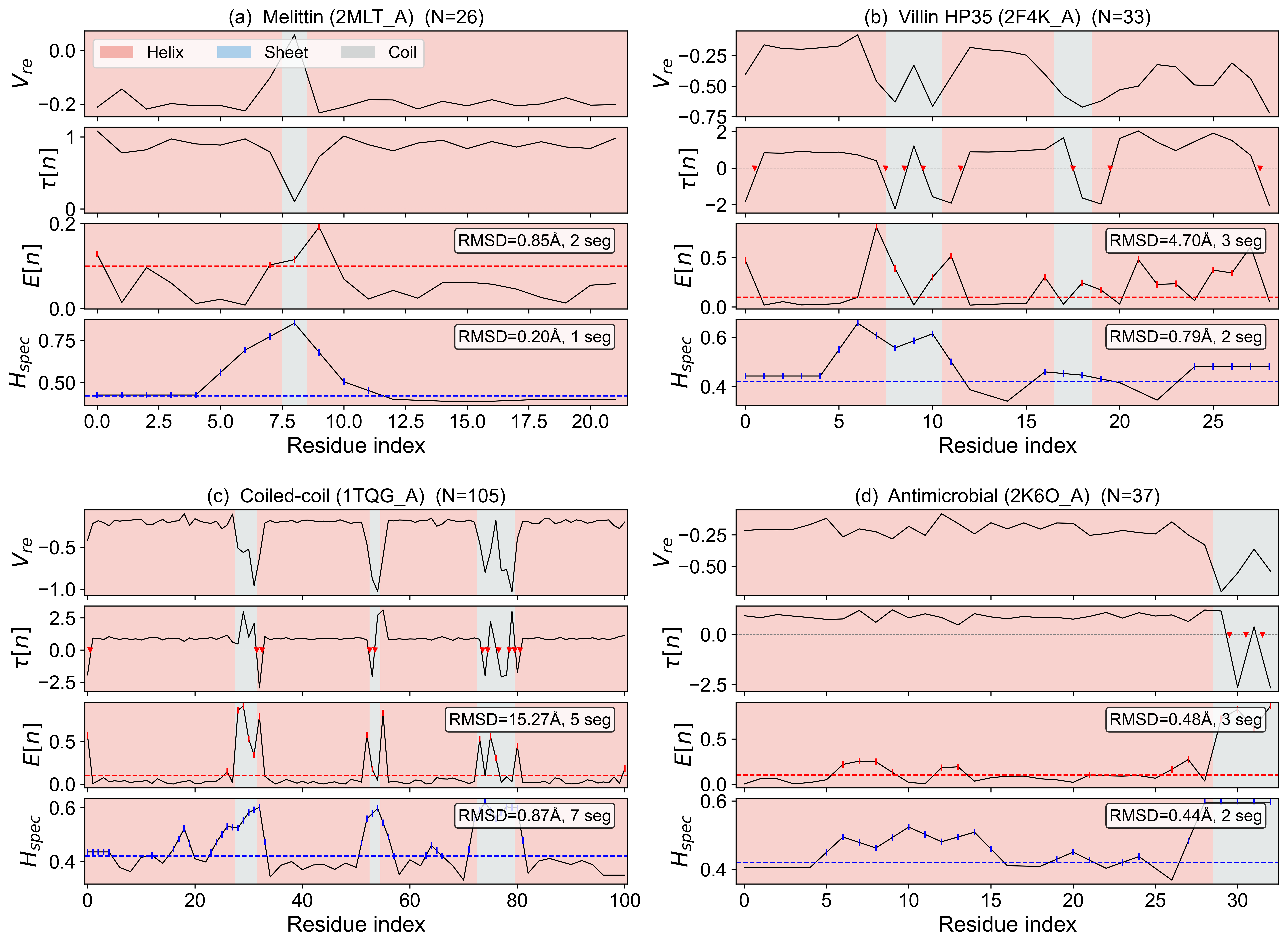}
		\caption{Macroscopic 3D manifestation of spatial-spectral trade-offs in helical peptide reconstruction. The four chains are illustrative examples of the dataset-wide reconstruction analysis reported here (72.9\% success for $E[n]$ over the full 251-chain set), selected to expose the boundary-precision versus defect-robustness trade-off. Each panel displays the local geometric signals ($V_{\mathrm{re}}[n]$ and $\tau[n]$ with sign-flip markers as red triangles) alongside the segmentation decisions made by the integrability probe $E[n]$ (red dashed, threshold $\theta = 0.10$) and the windowed spectral entropy $H_{\mathrm{spec}}$ (blue dashed, $\sigma^* = 1.5$, optimal threshold $\theta^* = 0.420$). Background shading indicates DSSP secondary structure (helix in red, sheet in blue, coil in gray). The resulting 3D RMSD and segment counts illustrate the conflict between boundary precision and robustness to local conformational defects. (a)~Clean boundary limit (2MLT\_A). In a highly ordered peptide with no torsion noise, $E[n]$ introduces a few spurious domain boundaries, but both methods successfully capture the core topology (RMSD $< 1.0$~\AA). (b) and (c)~Pointwise fragmentation and macroscopic recovery (2F4K\_A and 1TQG\_A). In sequences with localized geometric defects, the lattice-level sensitivity of $E[n]$ fragments the macroscopic continuous helix into disconnected segments (RMSD increases to $4.70$~\AA\ and $15.27$~\AA). The STFT window of $H_{\mathrm{spec}}$ averages over these defects to preserve the continuous near-integrable structure, recovering the global topology (RMSD drops to $0.79$~\AA\ and $0.87$~\AA). (d)~Marginal equilibrium (2K6O\_A). Both the local and macroscopic observables achieve comparable 3D segmentation on a moderately clean helix, highlighting that structural reconstruction relies on balancing these two complementary spatial scales.}
		\label{fig:cases}
	\end{figure*}

	However, case studies reveal a notable vulnerability of the pointwise limit that macroscopic spectral averaging can mitigate. For sequences containing localized geometric defects, such as isolated torsion sign-flips or local structural bends, the lattice-level sensitivity of $E[n]$ causes substantial domain fragmentation. For instance, in the coiled-coil 1TQG\_A (Fig.~\ref{fig:cases}c), local torsion fluctuations cause $E[n]$ to fragment a continuous macroscopic helix into five disconnected segments, increasing the 3D RMSD to $15.27$~\AA. The STFT window of $H_{\mathrm{spec}}$ averages over these localized defects, preserving the underlying continuous near-integrable structure and recovering the global helical topology (RMSD $= 0.87$~\AA). A similar recovery is observed in the villin headpiece (2F4K\_A, Fig.~\ref{fig:cases}b), where $H_{\mathrm{spec}}$ improves the RMSD by $3.91$~\AA\ by smoothing over localized noise.
	
	These 3D reconstructions illustrate that no single observable at a fixed spatial scale is globally optimal, given the resolution trade-off between pointwise and windowed spectral probes. Absolute boundary precision (requiring a pointwise local probe) and global robustness to local conformational defects (requiring a macroscopic averaging window) cannot be achieved simultaneously. We attempted an adaptive routing strategy, using per-chain torsion standard deviation $\sigma_\tau$ to explicitly choose between $E[n]$ and $H_{\mathrm{spec}}$ for each chain. This strategy did not improve the global success rate beyond 72.9\%, despite an oracle upper bound of 84.1\%, motivating the simultaneous multi-scale approach (Fig.~\ref{fig:concept}) rather than a binary per-chain selection.

	\begin{figure*}[!t]
		\centering
		\includegraphics[width=\textwidth]{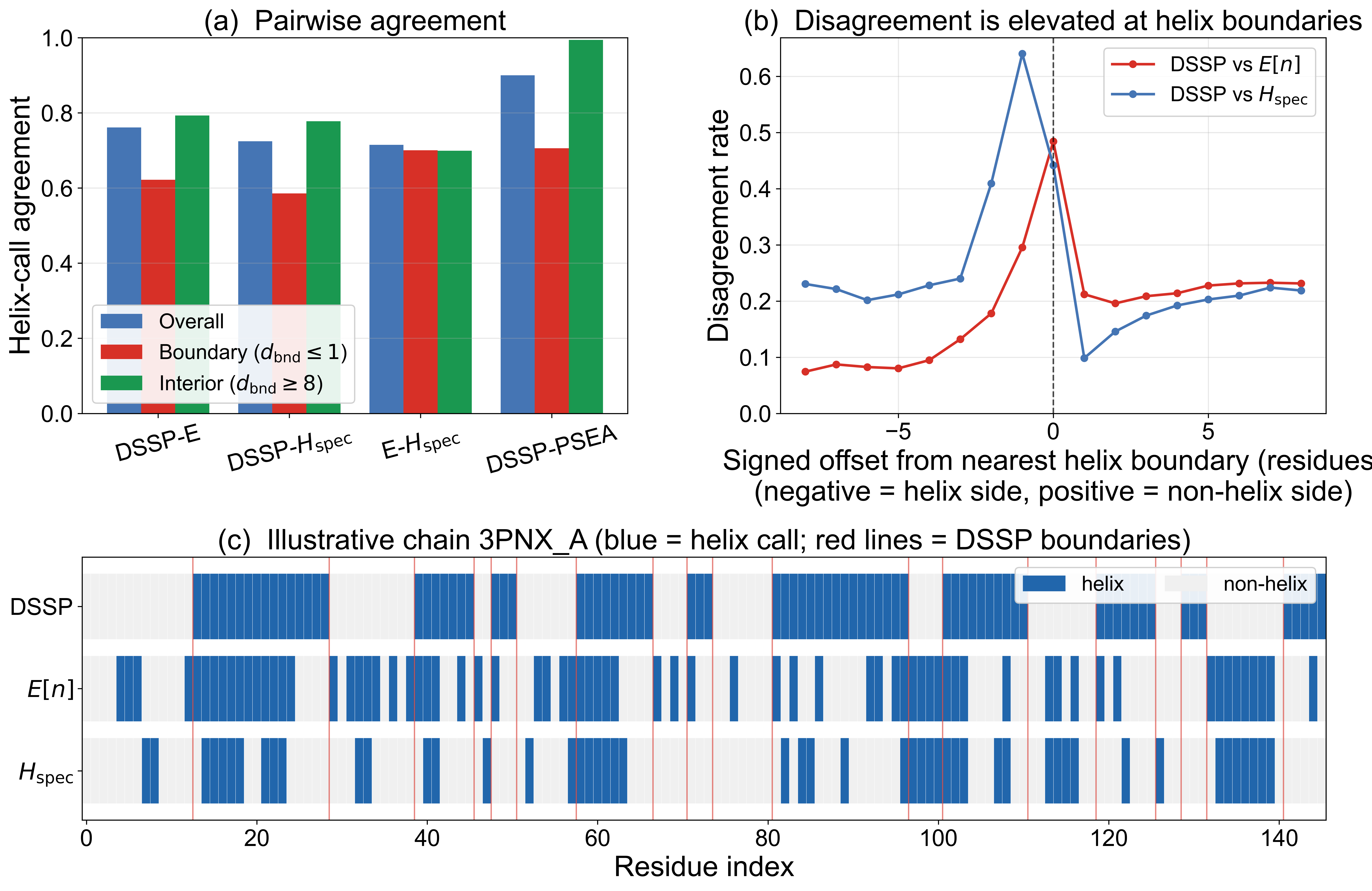}
		\caption{Structural-assignment disagreement concentrates at helix--coil boundaries across the full dataset (320\,453 residues, 1\,986 chains). (a)~Pairwise helix-call agreement for four assignment methods, stratified into overall, boundary ($d_{\mathrm{bnd}} \leq 1$), and interior ($d_{\mathrm{bnd}} \geq 8$) residues. Agreement between DSSP and each geometric probe ($E[n]$, $H_{\mathrm{spec}}$) is markedly lower at boundaries than in the interior, whereas the two probes agree with each other at a comparable rate in both regimes. The C$_\alpha$-only assigner P-SEA reproduces DSSP almost perfectly in the interior, reflecting its calibration as a DSSP-style classifier, yet still loses agreement at boundaries. (b)~Disagreement rate against signed offset from the nearest helix boundary (negative on the helical side, positive on the non-helical side). Disagreement peaks at the boundary and decays asymmetrically, approaching zero a few residues into the helical side while settling onto a method-dependent floor near 0.20 on the non-helical side, rather than forming a symmetric spike that decays on both sides. (c)~An illustrative single chain (3PNX\_A) showing the DSSP, $E[n]$, and $H_{\mathrm{spec}}$ helix calls as parallel tracks (blue denotes a helix call), with DSSP boundaries marked by vertical red lines; the three tracks coincide within helical interiors and diverge within one to two residues of each boundary.}
		\label{fig:disagreement}
	\end{figure*}

	These single-chain reconstructions visualize a dataset-wide trade-off. The reconstruction success rate (72.9\% for $E[n]$ across the full 251-chain helical set) and the boundary-resolved separability (Fig.~\ref{fig:sigma_bnd}) are both measured over the entire dataset, and the four chains of Fig.~\ref{fig:cases} show how this trade-off manifests in individual structures. To test whether the boundary ambiguity is specific to the present geometric probes or extends to other C$_\alpha$ geometry-based assignment methods, we compared four helix-versus-rest calls on all 320\,453 aligned residues, namely the hydrogen-bond reference DSSP, the pointwise integrability residual $E[n]$, the windowed spectral entropy $H_{\mathrm{spec}}$, and the C$_\alpha$-only assigner P-SEA, stratified by distance $d_{\mathrm{bnd}}$ to the nearest boundary (Fig.~\ref{fig:disagreement}). Consistent with the separability collapse in Fig.~\ref{fig:sigma_bnd}, agreement between DSSP and either geometric probe is far lower at boundary residues ($d_{\mathrm{bnd}} \leq 1$) than in the interior ($d_{\mathrm{bnd}} \geq 8$), with values of 0.622 against 0.793 for $E[n]$ and 0.586 against 0.778 for $H_{\mathrm{spec}}$. Crucially, the same localization holds for methods unrelated to our probes. The DSSP-calibrated P-SEA reproduces DSSP almost perfectly in the interior (0.995) yet agrees with it on only 0.706 of boundary residues, and three-way agreement among DSSP, $E[n]$, and $H_{\mathrm{spec}}$ falls to 0.454 at boundaries against 0.635 in the interior. The two geometric probes, by contrast, agree with each other at a rate that is essentially flat across layers (0.701 at boundaries versus 0.699 in the interior), because both inherit the same boundary-smearing limitation rather than disagreeing preferentially at the interface. The roughly 0.70 agreement level even in helix interiors reflects the intrinsic smoothing of the windowed probe over its $6\sigma \approx 9$-residue support, which blurs the isolated interior defects that the pointwise $E[n]$ resolves at the lattice limit, rather than any boundary effect. Even P-SEA, whose C$_\alpha$ distance rules were tuned to reproduce DSSP-style assignments, loses agreement with the hydrogen-bond reference at the boundary; the effect is therefore a general feature of C$_\alpha$ geometry-based helix assignment rather than a peculiarity of either spectral probe. The boundary is therefore the regime where C$_\alpha$ backbone geometry separates the two phases least cleanly, whether that geometry is read pointwise, through a spectral window, or through fixed distance rules, consistent with the near-chance boundary separability of Fig.~\ref{fig:sigma_bnd} and with the limited resolving power of the geometric probes raised in Sec.~\ref{sec:intro}.

	\section{Discussion \& Conclusion}
	\label{sec:discussion}
	
	The spectral contrast between DC-dominated helical segments and broadband coil regions, established quantitatively in Secs.~\ref{sec:order_disorder} and \ref{sec:gabor_limit}, provides a frequency-domain rationale for the empirical success of the pointwise integrability residual $E[n]$ \cite{wang2026structural, wang2026piecewise}. Because $E[n]$ measures the local violation of the DNLS dispersion relation, it vanishes within the near-constant helical plateau and responds sharply to boundary discontinuities where torsion non-uniformity disrupts integrability. Combined with the single-step sharpness of helix--coil transitions, this contrast renders any finite-width spectral window susceptible to boundary contamination, giving rise to an intrinsic spatial-spectral trade-off between boundary precision and robustness to local conformational defects.
	
	This intrinsic trade-off implies that relying solely on the pointwise local limit is topologically fragile \cite{ferrari2019knots}. Isolated torsion defects within an otherwise continuous helix trigger fragmentation of the macroscopic domain. Conversely, widening the observation window to absorb such local noise necessarily blurs the single-residue-step phase boundaries. An ideal discrete nonlinear Schr\"{o}dinger (DNLS) soliton propagates without dispersing its spectral content, producing minimal spectral entropy; real helices approximate this integrable limit. The local order parameter ($E[n]$) detects the rupture of this near-integrability at boundaries and defects, while the macroscopic spectral probe ($R_{\mathrm{LF}}$) measures the preservation of the DC-dominated spectral core within helical interiors. Their combination captures complementary structural information not fully accessible to either observable alone, raising the phase-state separability from AUC 0.783 to 0.815.
	
	The single-step sharpness of helix--coil transitions (median width $w = 0.145$ residues across 19,148 fitted boundaries) requires careful physical interpretation. On the discrete peptide lattice, this value means that the sigmoid transition in spectral entropy is effectively complete between one $C_\alpha$ position and the next, with no intermediate partial-helical state; the fitted continuous width should therefore be understood as a measure of the abruptness of the underlying discrete jump rather than as a literal sub-residue spatial extent. This step-like character finds a natural thermodynamic counterpart in the Zimm--Bragg helix--coil model \cite{zimm1959theory, flack2023generalized}, a foundational framework in polymer and soft matter physics, where the small nucleation parameter ($\sigma_{\mathrm{ZB}} \ll 1$) compresses the coexistence region to one or two residues. The single-step geometric sharpness observed here provides an independent, purely kinematic signature consistent with this high degree of thermodynamic cooperativity. While the Hasimoto effective potential is a purely kinematic construct derived from static coordinates, this single-step sharpness serves as a structural footprint of the underlying energetic cooperativity that governs helix nucleation and propagation. The same helix--coil transition has independently been characterized by recent calorimetry, which reports a per-residue backbone conformational entropy change of $\Delta S_{\mathrm{BB}} = 5.2 \pm 0.3$~cal\,mol$^{-1}$\,K$^{-1}$ \cite{zavrtanik2026backbone}; this finite, well-resolved entropy difference and the single-step geometric sharpness observed here are two independent signatures, thermodynamic and kinematic respectively, of the same underlying two-state cooperativity. The directional asymmetry quantified in Sec.~\ref{sec:boundary_sharpness} (helix exits being sharper than helix entries) is likewise consistent with the cooperative thermodynamics of the Zimm--Bragg framework, in which the free-energy landscape for helix propagation differs from that for nucleation \cite{lopez2025orientation}, and it is complementary to the distinct terminal geometries documented in the helix capping literature \cite{segura2012caps, viguera1995experimental}.
		
	This geometric abruptness has a direct observational consequence through the Gabor limit \cite{gabor1946electrical}. Widening the STFT observation window improves spectral resolution but progressively blurs the single-residue-step boundary. Because the geometric transition approximates a single-step function, any window expansion beyond the minimal scale introduces boundary contamination, and the monotonic decay of AUC with increasing $\sigma$ is an empirical manifestation of this trade-off. The failure of the adaptive per-chain routing strategy to surpass the pointwise baseline further illustrates that the conflict between boundary precision and robustness to local conformational defects operates at the residue level, not the chain level. This trade-off is a general constraint of windowed spectral analysis applied to structural boundary detection, not a limitation specific to the present method. It does not, by itself, account for the well-known observation that existing secondary structure assignment algorithms, including DSSP, STRIDE, and their successors, systematically disagree at helix--coil boundaries by one to two residues \cite{martin2005protein, kannan2024comprehensive}; that inter-algorithm discrepancy arises primarily from their differing structural definitions and energy-cutoff choices. What our spectral analysis adds is that windowed probes are subject to a further, method-intrinsic resolution limit at these same boundaries, whereas the pointwise probe is not. The present framework is therefore not intended to supersede these established assignment tools in speed or coverage, but to characterize the frequency-domain structure of the helix--coil boundary and to quantify how the resolution of windowed versus pointwise spectral probes limits its measurement. The AUC values reported here accordingly quantify the geometric distinguishability of the backbone attainable by each probe. Residue-level multi-resolution fusion or nonlinear probe combinations may partially alleviate it and remain directions for future work.

	A methodological limitation is the reliance on X-ray crystallographic structures, which represent low-temperature or cryogenic single-conformer snapshots. Under physiological conditions, thermal fluctuations induce helix-end fraying and breathing motions \cite{zimm1959theory} that would broaden the helix--coil interface; the median width of $0.145$ residues is therefore a lower bound characteristic of the crystallographic ensemble. The high-frequency geometric noise that triggers domain fragmentation by $E[n]$ in the case studies (Sec.~\ref{sec:gabor_limit}) represents genuine local conformational features, including proline-induced kinks, coiled-coil superhelical distortions, and backbone flexibility frozen in the crystal lattice, rather than experimental coordinate errors, which at 2.0~\AA\ resolution perturb $\kappa$ by only $\sim 0.01$~rad, far below the $\sim 0.5$~rad torsion jumps that activate $E[n]$. The STFT smoothing thus performs a physically meaningful operation by averaging over local conformational heterogeneity to recover the macroscopic helical topology.

	The Hasimoto map encodes exclusively the $C_\alpha$ backbone kinematics (bond-angle curvature and dihedral torsion) and is by construction blind to side-chain packing, hydrophobic interactions, and solvent effects, all of which contribute substantially to folding thermodynamics. The spectral signatures reported here thus characterize the backbone geometric component of secondary structure and do not capture the full energetic landscape. The standardized linear combination used to fuse the local and macroscopic observables is the simplest possible combination strategy; nonlinear methods or locally adaptive weighting schemes may yield further improvements, though at the cost of additional free parameters. The discrete Frenet frame \cite{danielsson2010gauge} uses forward differences $\Delta\mathbf{r}[n] = \mathbf{r}[n{+}1] - \mathbf{r}[n]$. Repeating the full pipeline with centered differences $(\mathbf{r}[n{+}1] - \mathbf{r}[n{-}1])/2$ preserves all qualitative conclusions, with the spectral entropy ordering $\langle H_{\mathrm{spec}}\rangle_{\mathrm{helix}} < \langle H_{\mathrm{spec}}\rangle_{\mathrm{coil}}$, the integrability residual ordering $\langle E\rangle_{\mathrm{helix}} < \langle E\rangle_{\mathrm{coil}}$, and the mean helical $E$ value ($0.132$ vs.\ $0.133$) essentially unchanged; the centered-difference AUC values are numerically higher because the smoothed tangent amplifies the spectral contrast between near-integrable helices and disordered coils, confirming that the forward-difference choice is conservative and the conclusions are not artifacts of the discretization (Table~\ref{tab:frenet_robustness}). The boundary analysis relies on DSSP \cite{kabsch1983dictionary} assignments as reference annotations, and DSSP itself employs a fixed hydrogen-bond energy threshold that introduces its own boundary ambiguity. Alternative assignment algorithms (e.g., STRIDE \cite{frishman1995knowledge}) or torsion-angle clustering approaches \cite{kannan2024comprehensive} may shift individual boundary positions by one to two residues, although the statistical conclusions drawn from $>$19,000 boundaries are unlikely to be qualitatively affected.

	The multi-scale framework also has an intrinsic minimum chain-length requirement imposed by the STFT window. At $\sigma = 1.5$, the Gaussian window spans $M = 9$ residues; combined with the Frenet-frame end trimming ($\pm 2$ residues), a chain of $N$ C$_\alpha$ atoms yields only $N - 12$ valid $H_{\mathrm{spec}}$ values. Chains shorter than approximately 20 residues therefore provide too few spectral estimates for the windowed probe to characterize helical interiors reliably, and for chains below 13 residues no $H_{\mathrm{spec}}$ value can be computed at all. In this short-peptide regime, relevant to de novo peptide design and antimicrobial peptide engineering, the pointwise integrability residual $E[n]$, which requires only $N \geq 7$ and carries no window constraint, remains the sole applicable geometric descriptor. For intrinsically disordered proteins (IDPs), the framework correctly identifies the entire chain as broadband and high-entropy, but the absence of DC-dominated helical segments means that the macroscopic probe $R_{\mathrm{LF}}$ provides no complementary information, reducing the multi-scale framework to the single-observable $E[n]$ limit.
	
	The static-snapshot limitation points directly to a natural extension of this work. Because the Hasimoto map is defined at each instantaneous backbone configuration, it can be applied frame-by-frame to molecular dynamics (MD) trajectories, yielding a time-dependent effective potential $V_{\mathrm{re}}[n,t]$. A joint spatial-temporal spectral decomposition of this field would enable tracking of helix-end breathing, the propagation of conformational perturbations along the backbone, and the spectral signatures of folding or allosteric transitions during their temporal evolution, building on systematic characterizations of protein family conformational diversity \cite{lombard2024explaining}. More broadly, the spectral entropy $H_{\mathrm{spec}}$ provides a sequence-agnostic, purely geometric measure of local conformational variability on the 1D backbone lattice, complementing the graph-derived local Shannon entropy recently introduced for conformational substates in molecular dynamics \cite{senet2026local} and directly connecting the geometric order parameter to the soft-matter concept of conformational entropy density. High-entropy broadband regions tend to correspond to conformationally flexible segments such as hinges, loops, and intrinsically disordered regions \cite{wright1999intrinsically} that have been associated with allosteric communication and protein--protein interactions \cite{wankowicz2025making}. For IDPs in particular, the static broadband signature correctly identifies the absence of stable helical order but cannot resolve transient local structuring events such as helical nucleation, nascent $\beta$-hairpin formation, and the emergence of cryptic binding pockets \cite{moses2024structural, qin2025current} that govern molecular recognition and function in these entropy-dominated systems. The time-resolved extension to MD trajectories is therefore especially relevant for detecting when and where fleeting geometric order nucleates within the disordered ensemble. This question carries direct therapeutic significance, given that transient pre-structured conformations in IDPs have recently been shown to serve as druggable targets through conformational selection \cite{bogin2025drugging}. Unlike sequence-based spectral methods that transform one-dimensional physicochemical indices into frequency representations \cite{veljkovic1972simple, cosic2002macromolecular, wang2025signal}, the present framework operates directly on the three-dimensional differential geometry of the folded backbone, offering a complementary structural perspective. Whether this geometric spectral fingerprint can quantitatively predict functional flexibility or allosteric pathways, potentially serving as an input feature for flexibility predictors \cite{kouba2024learning}, remains to be validated against experimental dynamics data and larger-scale functional annotations.

	A further limitation of the present C$_\alpha$-only Hasimoto map is that it is by construction blind to the peptide-plane degrees of freedom, specifically the N--C$_\alpha$--C$'$=O dihedral constraints and amide hydrogen positions, that directly mediate backbone hydrogen bonding. Extending the map to incorporate this full peptide-plane geometry would provide access to the vibrational coordinates that underpin the Davydov soliton model \cite{davydov1977solitons} of nonlinear energy transfer along $\alpha$-helices, potentially bridging the present kinematic description with a Hamiltonian treatment of backbone dynamics that explicitly accounts for the N--H$\cdots$O=C hydrogen-bond network. More generally, because the Hasimoto map operates on any discrete space curve regardless of chemical identity, the spectral-entropy framework is in principle applicable to any soft matter system whose backbone geometry undergoes cooperative order--disorder transitions. Natural candidates include DNA supercoiling, where plectoneme nucleation represents an analogous symmetry-breaking event on a semiflexible polymer \cite{daniels2011nucleation, junier2023dna}, and the helix--coil equilibria of synthetic polypeptoids \cite{kirshenbaum1998sequence, sanborn2002extreme}, where the absence of backbone hydrogen-bond donors produces a distinct cooperativity landscape. The geometric-spectral mapping developed here thus offers a unified framework for characterizing structural phase boundaries across the broader class of semiflexible biopolymers and synthetic macromolecules that constitute canonical soft condensed matter systems.
	
	For C$_\alpha$ geometry-based assignments the persistent boundary ambiguity substantially reflects the limited resolving power of the geometry itself, since $E[n]$, $H_{\mathrm{spec}}$, and DSSP-calibrated P-SEA all lose agreement with the DSSP hydrogen-bond reference at the boundary while largely agreeing in the interior, operating alongside the definitional differences that drive the discrepancies among hydrogen-bond algorithms. In summary, this study establishes the helix--coil boundary of protein backbones as a model system for investigating observational limits in soft matter structural characterization. The single-step sharpness of helix--coil transitions (median $w = 0.145$ residues) provides independent kinematic evidence for the high thermodynamic cooperativity described by the Zimm--Bragg model \cite{zimm1959theory}, and the accompanying resolution trade-off explains both the strengths and the vulnerabilities of the windowed spectral probe relative to the pointwise probe. The multi-scale characterization framework, combining a local order parameter with a macroscopic spectral probe, offers a principled response to this physical constraint and improves structural-state separability beyond either observable alone. Applying a spectral analysis of C$_\alpha$ backbone geometry across a large structural dataset, we show that helix--coil boundaries are abrupt and directionally asymmetric, with helix exits sharper than entries. Windowed spectral probes carry an intrinsic Gabor resolution limit at these boundaries that pointwise probes avoid; this offers a unified geometric perspective applicable to cooperative transitions in biopolymers and synthetic soft matter systems.
	
	\section*{Acknowledgments}
	We thank Prof. Yan-Hong Qin (Xinjiang University) for inspiration and foundational guidance that motivated this work. This work was carried out in part using computing resources at the Computing and Data Center of Xinjiang University.

	\section*{Ethics declarations}
	\subsection*{Conflict of interest}
	The authors declare no conflict of interest.

	\section*{CRediT authorship contribution statement}
	\textbf{Yiquan Wang:} Conceptualization, Methodology, Software, Validation, Formal analysis, Investigation, Data curation, Writing -- original draft, Writing -- review \& editing, Visualization.
	
	\section*{Data and Code Availability}
	The source code for this study is available on GitHub at \url{https://github.com/wyqmath/discrete_hasimoto_protein}.

	\bibliographystyle{unsrt}
	\bibliography{main}

	\setcounter{figure}{0}
	\renewcommand{\thefigure}{S\arabic{figure}}
	\setcounter{table}{0}
	\renewcommand{\thetable}{S\arabic{table}}
	\clearpage

	\begin{figure*}[t!]
		\centering
		\includegraphics[width=1\linewidth]{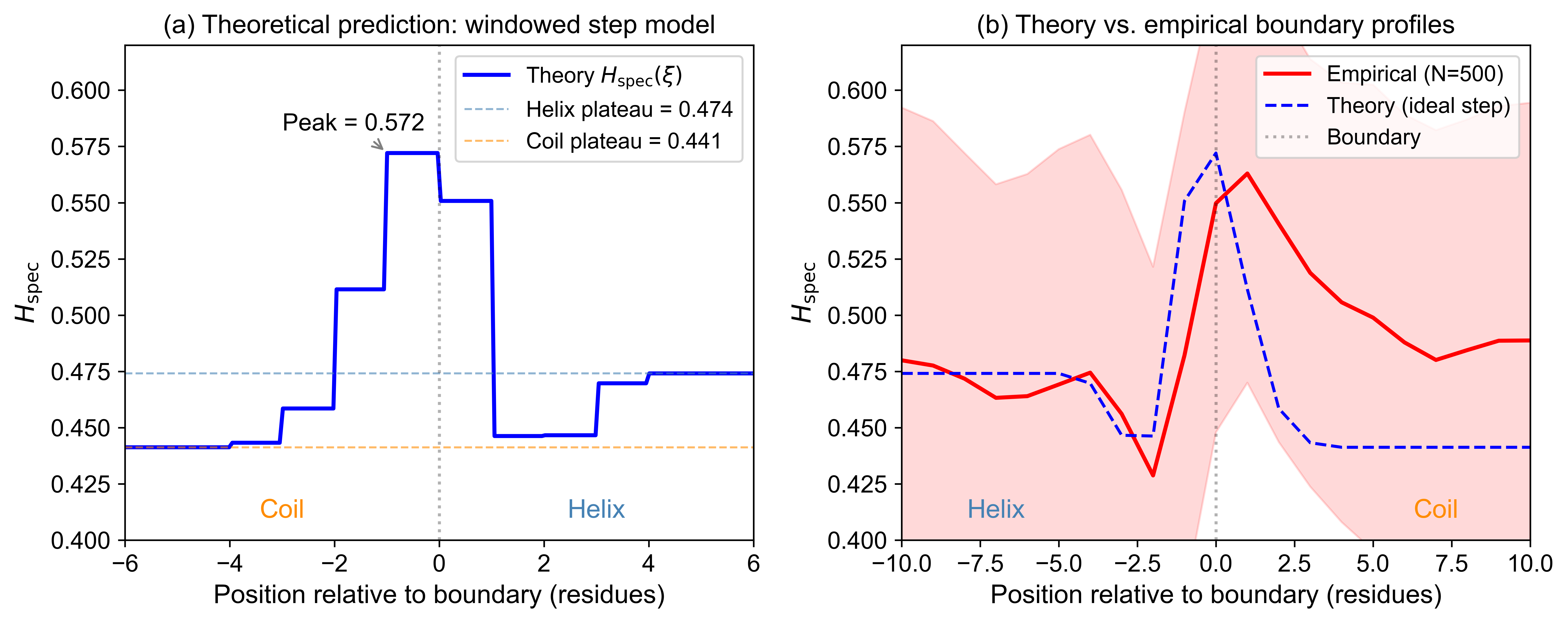}
		\caption{\label{fig:S1}Numerical validation of the theoretical boundary resolution limit (Eq.~\ref{eq:Pk_boundary}). (a)~Theoretical prediction of $H_{\mathrm{spec}}(\xi)$ as a function of window position relative to the helix--coil boundary. The model uses a windowed step function for the mean signal combined with measured fluctuation spectral shapes. The entropy peaks near the boundary owing to spectral leakage from the mean discontinuity, with the helix plateau (0.474) slightly exceeding the coil plateau (0.441) because the larger coil DC level ($|\mu_C|^2 = 6|V_0|^2$) concentrates more power at zero frequency. (b)~Comparison of theoretical prediction (blue dashed) with the empirical mean boundary profile (red solid, averaged over 500 helix--coil transitions). The theory accurately reproduces the peak height (0.572 vs.\ 0.563), peak position (+1~residue into coil), and FWHM (2.0~residues). The shaded region indicates $\pm 1$ standard deviation across empirical profiles. Monte Carlo validation with colored noise (5000 trials) confirms the analytical model to within 0.02 (Jensen inequality bias).}
	\end{figure*}

	\clearpage

	\begin{table*}[t!]
		\caption{\label{tab:frenet_robustness}Robustness of discrimination metrics to the choice of discrete derivative in the Frenet frame. Forward difference: $\mathbf{t}[n] = (\mathbf{r}[n{+}1] - \mathbf{r}[n])/|\cdot|$; centered difference: $\mathbf{t}[n] = (\mathbf{r}[n{+}1] - \mathbf{r}[n{-}1])/|\cdot|$. All values computed on the full dataset (1,986 chains, $>$300k residues). The qualitative ordering $\langle\cdot\rangle_{\mathrm{helix}} < \langle\cdot\rangle_{\mathrm{coil}}$ is preserved for both $H_{\mathrm{spec}}$ and $E[n]$. The higher AUC with centered differences reflects the intrinsic spatial smoothing of the symmetric derivative, which amplifies coil-region fluctuations while leaving the near-uniform helical geometry essentially unchanged ($\langle E\rangle_{\mathrm{helix}}$: 0.132 vs.\ 0.133).}
		\centering
		\begin{tabular}{lccc}
			\toprule
			Metric & Forward difference & Centered difference & $\Delta$ \\
			\midrule
			$H_{\mathrm{spec}}$ AUC & 0.715 & 0.858 & $+$0.144 \\
			$E[n]$ AUC & 0.783 & 0.866 & $+$0.082 \\
			$\langle H_{\mathrm{spec}}\rangle_{\mathrm{helix}}$ & 0.440 & 0.492 & $+$0.052 \\
			$\langle H_{\mathrm{spec}}\rangle_{\mathrm{coil}}$ & 0.513 & 0.737 & $+$0.224 \\
			$\langle E\rangle_{\mathrm{helix}}$ & 0.133 & 0.132 & $-$0.001 \\
			$\langle E\rangle_{\mathrm{coil}}$ & 0.386 & 0.656 & $+$0.270 \\
			\bottomrule
		\end{tabular}
	\end{table*}

\end{document}